\documentclass[a4paper,11pt]{article}
\pdfoutput=1 % if your are submitting a pdflatex (i.e. if you have
\usepackage{jcappub} % for details on the use of the package, please
\usepackage[normalem]{ulem}

\usepackage{amsmath,amssymb}
\usepackage{epsfig}  
\usepackage{graphicx}   
\usepackage{slashed}             
\usepackage{url}
\usepackage{color}
\usepackage{multirow}
\usepackage[dvipsnames]{xcolor}
\usepackage{letltxmacro}
\LetLtxMacro{\oldcite}{\cite}
\renewcommand{\cite}[1]{\mbox{\oldcite{#1}}}

\allowdisplaybreaks

\bibpunct{[}{]}{,}{n}{}{,}

\begin{document}

\begin{flushleft}
DESY 20-121, LAPTH-033/20
\end{flushleft}

\author[a,b]{Gautham A. P.,}% Adamane Pallathadka,}
\affiliation[a]{Indian Institute of Technology Madras (IITM), Chennai, India}
\affiliation[b]{Univ.~Grenoble Alpes, USMB, CNRS, LAPTh, F-74000 Annecy, France}

\author[b]{Francesca Calore,}

\author[c,d]{Pierluca Carenza,}
\affiliation[c]{Dipartimento Interateneo di Fisica ``Michelangelo Merlin'', Via Amendola 173, 70126 Bari, Italy}
\affiliation[d]{Istituto Nazionale di Fisica Nucleare - Sezione di Bari,
Via Orabona 4, 70126 Bari, Italy}

\author[e]{Maurizio Giannotti,}
\affiliation[e]{Physical Sciences, Barry University, 11300 NE 2nd Ave., Miami Shores, FL 33161, USA}

\author[f]{Dieter Horns,}
\affiliation[f]{Institut f\"{u}r Experimentalphysik, Universit\"at Hamburg,
Hamburg,Germany}

\author[f]{Julian Kuhlmann,}

\author[g]{Jhilik Majumdar,}
\affiliation[g]{Deutsches Elektronen-Synchrotron DESY, D-15738 Zeuthen, Germany}

\author[c,d]{Alessandro Mirizzi,}

\author[h]{Andreas Ringwald,}
\affiliation[h]{Deutsches Elektronen-Synchrotron DESY, Notkestra\ss e 85, D-22607 Hamburg, Germany}

\author[h,i]{Anton Sokolov,}
\affiliation[i]{Institute for Nuclear Research of the Russian Academy of Sciences, 60th October Anniversary pr. 7a, 117312 Moscow, Russia}

\author[f]{Franziska Stief,}

\author[f]{Qixin Yu}

\emailAdd{ph16b002@smail.iitm.ac.in, calore@lapth.cnrs.fr, pierluca.carenza@ba.infn.it, MGiannotti@barry.edu, dieter.horns@physik.uni-hamburg.de, julian.kuhlmann@physik.uni-hamburg.de, jhilik.majumdar@desy.de, alessandro.mirizzi@ba.infn.it,
andreas.ringwald@desy.de, anton.sokolov@desy.de, franziska.stief@desy.de, qixin.yu@desy.de}

\title{Reconciling hints on axion-like-particles from high-energy gamma rays with stellar bounds}

\date{\today}
%=============================================================================

\abstract{
It has been recently claimed by two different groups that the spectral modulation observed in gamma rays
from Galactic pulsars and supernova remnants can be due to conversion of photons into ultra-light axion-like-particles (ALPs) in large-scale Galactic  magnetic fields. 
While we show the required best-fit photon-ALP coupling, $g_{a\gamma} \sim 2 \times 10^{-10}$ GeV${}^{-1}$, to be consistent with constraints from observations of photon-ALPs
mixing in vacuum, this
is in conflict with 
other bounds, specifically from the CAST solar axion limit, from the helium-burning lifetime in globular clusters, and from  the non-observations of gamma rays in coincidence
with SN 1987A. In order to reconcile these different results, we propose that environmental effects in matter would suppress the ALP production in dense astrophysical plasma, allowing to relax previous bounds and make them compatible with photon-ALP conversions in the low-density Galactic medium.  If this explanation is correct, the claimed ALP signal  would be on the reach of next-generations laboratory experiments such as ALPS II. 
}

\maketitle

\section{Introduction}
Axion-like particles (ALPs) are ultra-light pseudo-scalar bosons $a$ with a two-photon vertex $a \gamma \gamma$,
predicted by several extensions of the Standard Model (for a review, see~\cite{Jaeckel:2010ni,DiLuzio:2020wdo}).
 The two-photon coupling allows the conversion of ALPs into photons, $a\leftrightarrow\gamma$, in external electric or magnetic fields.
In stars, this leads to  the Primakoff process that allows for the production of low mass
ALPs in the microscopic electric fields of nuclei and electrons. 
An ALP flux would then cause a novel source of energy-loss in stars, altering their evolution.
The strongest bound comes from the helium-burning stars in globular clusters, giving $g_{a\gamma} <  6.6 \times 10^{-11}$~GeV${}^{-1}$ for 
$m_a \lesssim 1$~keV~\cite{Ayala:2014pea}. In
the other case of a macroscopic field, usually a large-scale magnetic field,
the momentum transfer is small, the interaction is coherent over a large
distance, and the conversion is best viewed as an ALP-photon oscillation
phenomenon in analogy to neutrino flavor oscillations. 
This  effect is exploited to search for generic ALPs in light-shining-through-the-wall experiments (see e.g.~the ALPS~\cite{Ehret:2010mh} and
 OSQAR~\cite{Ballou:2015cka} experiments), for solar ALPs (see e.g.~the CAST experiment~\cite{Arik:2011rx,Arik:2013nya}) and for ALP dark 
 matter~\cite{Arias:2012az} in micro-wave cavity experiments (see e.g.~the ADMX experiment~\cite{Braine:2019fqb}). 
Discarding the narrow band probed by ADMX, the best experimental bound on the photon-ALP coupling is $g_{a\gamma}\lesssim g_{\rm CAST}=6.6 \times 10^{-11}$~GeV$^{-1}$ obtained by the CAST experiment for 
$m_a\lesssim 0.02$~eV~\cite{Anastassopoulos:2017ftl}. 
See~\cite{Irastorza:2018dyq,Sikivie:2020zpn} 
for a complete and updated overview of current and future plans for ALP searches.

 Due to the $a \gamma \gamma$ coupling, ultra-light ALPs can also play an important role  in astrophysical observations. 
  In particular, about a decade ago it was realized that conversions of 
 very high-energy gamma rays into ALPs in cosmic magnetic fields, would lead to peculiar
 signatures in the photon spectra from distant sources, allowing to probe a region of the ALP parameter
 space untouched by current experiments~\cite{Meyer:2013pny}. In particular, for $m_a \lesssim 10^{-7}$~eV and 
 $g_{\rm a\gamma}\gtrsim 10^{-11}$~GeV$^{-1}$, photon-ALP conversions in large-scale magnetic fields
 might  explain an anomalous spectral hardening found in the very high-energy gamma-ray spectra~\cite{DeAngelis:2011id}. 
 Another peculiar signature of  photon mixing with ALPs is an energy-dependent modulation of high-energy gamma-ray~\cite{Meyer:2014epa} or X-ray~\cite{Dessert:2020lil}
 spectra.
 
A search for such an effect has been recently performed in~\cite{Majumdar:2018sbv}, 
where the authors  analyzed the data recorded with the {\it Fermi}-LAT from bright Galactic pulsars (PSRs), detecting significant spectral features consistent with ALP-photon oscillation. 
 This hint was independently confirmed in~\cite{Xia:2018xbt} by analyzing {\it Fermi}-LAT data from bright supernova remnants. 
The existence of ALPs with such parameters 
is in tension with the previously mentioned
 astrophysical bound from globular cluster stars and with the direct bound on solar axions from CAST  (see also \cite{Choi:2018mvk} for an analogous approach to evade stellar bounds). 
 Given the robustness of these latter bounds, one can attribute the ALP hint to some unrecognized systematic effect. Still, without any simple alternative
 explanation, one can also investigate if there is a possibility to reconcile the two, apparently contradicting, results. This is the goal of the present work.
 
The apparent tension between a bound and positive hint for ALPs is reminiscent of what happened in 2005 when the PVLAS 
 collaboration reported  the observation of a rotation of the polarization plane of a laser propagating through a transverse magnetic 
 field~\cite{Zavattini:2005tm}. Interpreted 
 as an ALP induced effect, this would correspond to a particle with $m_a \sim 1$ meV and  $g_{a\gamma} =  10^{-6}$ GeV${}^{-1}$. 
 Obviously, this signal was in a strong tension with CAST and globular cluster bounds~\cite{Raffelt:2005mt}. 
 Indeed,  a few years later the claim was retracted by the collaboration~\cite{Zavattini:2007ee}. 
 However, the controversy led to an intense investigation on possible models to reconcile the two results. 
 These models can still represent an intriguing possibility to reconcile the ALP PSR hint with the 
 CAST and globular cluster  bounds. Remarkably, the tension between these two results is much milder than the one related to the PVLAS claim.
 For our purpose, we find particularly interesting the models proposed in~\cite{Jaeckel:2006xm}, where it was speculated 
 that the coupling and the mass of an ALP may depend on environmental conditions such as the temperature and matter density. 
 Within this framework one can achieve a sizable suppression of the ALP production in the high-density stellar plasma, for example in the Sun or in globular cluster stars, 
but also a significant photon-ALP mixing in the low-density Galactic medium, reconciling the 
 apparent tension. In what follows, we will show how this scenario works.

 In Sec.~\ref{sec:VHE} we present the current status of hints and bounds on ALPs from high-energy gamma rays.
In particular, we test the robustness of the PSR signal region of Ref.~\cite{Majumdar:2018sbv} against some systematic uncertainty, and we present updated constraints from NGC 1275 with {\it Fermi}-LAT and from PKS 2155-304 with H.E.S.S.~data.
   In Sec.~\ref{sec:dynamical_suppression}, we show how to dynamically  suppress the solar ALP flux via environmental dependence of the ALP-photon coupling.
  In Sec.~\ref{sec:phenomenology} we discuss the phenomenological consequences of the ALP-photon coupling suppression 
  in relaxing the CAST and stellar bounds and make them compatible with the PSR hint. We also
      predict the expected signal from ALP-photon coupling like the one required to explain the PSR signal in a pure laboratory experiment, like ALPS II~\cite{Bahre:2013ywa}. 
     Finally, in Sec.~\ref{sec:conclusions} we discuss our results and we conclude.

  %%%%%%%%%%% 
  \section{ALPs and high-energy gamma rays}
  %%%%%%%%%%%
  \label{sec:VHE}
  
  %%%%%%%%%%% 
  \subsection{Hints for ALP-photon coupling}
  %%%%%%%%%%%

In the past few years, several, independent, groups have analyzed the gamma-ray spectra of high-energy sources, and, for some of those, unveiled a preference for the presence of ALP-photon conversion in the Galaxy.
%{observations of high-energy source spectra with gamma-ray telescopes have unveiled a preference for the presence of ALP-photon conversion in the Galaxy.}
Given the many analyses performed, we will offer here a review of what are current signal hints for ALP-photon conversion in the Galaxy, and, in the following section, what astrophysical bounds already exist which are based on the same environmental conditions.

Typically, searches towards Galactic objects have the advantage that they require to model only the conversion in the Galactic magnetic field. On the other hand, the strength of the ALP-photon conversion signal very much depends on the position of the source with respect to, for example, the Galactic spiral arms and so not all sources are optimal targets for this type of search.
  
In \cite{Majumdar:2018sbv}, the search for energy-dependent modulations in the gamma-ray spectra of six (bright and close by) PSRs detected by the {\it Fermi}-LAT telescope revealed a 4.6$\sigma$ preference for ALP-photon conversion in the large-scale Galactic magnetic field. The combined statistical analysis indicates as best-fit parameters $g_{a \gamma} = (2.3\pm0.4) \times 10^{-10}$ GeV${}^{-1}$ and $m_a = (3.6\pm 0.3)$ neV (statistical uncertainties only, systematic uncertainties are similar in magnitude).
Systematic and instrumental effects are unlikely to cause the spectral modulation as was demonstrated by analyzing the nearby Vela PSR, where modulations are expected to be very small.
In Sec.~\ref{sec:updatePSR}, we re-assess the PSR signal region of Ref.~\cite{Majumdar:2018sbv}, by fully taking into account distance and magnetic field uncertainties.
   
   In \cite{Xia:2019yud}, the GeV ({\it Fermi}-LAT) to TeV (H.E.S.S., MAGIC, VERITAS) spectra of three bright supernova remnants (IC443, W51C and W49B) were analyzed to look for ALP-photon conversion induced oscillations. Only a marginal signal ($\sim 3\sigma$) for ALP-photon conversion was found from IC443 at $m_a = 33.5$ neV and $g_{a\gamma} = 0.68 \times 10^{-10}$ GeV$^{-1}$, amending results from~\cite{Xia:2018xbt}. A lower significance was obtained for the other two sources. The authors carefully noticed that the slight preference for ALPs is mainly contributed by TeV data points and that, therefore, this result may be driven by a mismatch in the absolute energy calibration of low-energy ({\it Fermi}-LAT) and high-energy (H.E.S.S., MAGIC, VERITAS) data. 
   Given the absence of a clear preference for ALP-photon conversion,  Ref.~\cite{Xia:2019yud} sets limits in the ALPs parameter space through a combined statistical analysis (see also Sec.~\ref{sec:bounds}).  
   
   Finally, \cite{Liang:2018mqm} considered other Galactic sources also at TeV energies looking for ALPs oscillation effects in the spectra of a sample of ten supernova remnants and PSR wind nebulae. From the combined analysis, a 1.4$\sigma$ preference for ALPs emerged and upper bounds in the ALPs parameter space were set. 
   
Also TeV gamma-ray spectra of extragalactic sources bring us information about possible evidence of  ALP-photon conversion (see also \cite{Galanti:2015rda}). On the one hand, the excess in the cosmic infrared background at about 1 $\mu$m measured by the CIBER collaboration seems to suggest a significant attenuation in the spectra of TeV sources, whose photons interact with the infrared background during propagation to Earth.  
In \cite{Kohri:2017ljt}, this strong absorption not being seen in the {\it Fermi}-LAT and H.E.S.S. spectra of two high-energy sources ($z \sim 0.15$) was interpreted in terms of ALP-photon conversion for $m_a = 0.7 - 50$ neV and $g_{a \gamma} = 0.15 \times 10
   ^{-10} - 8.8 \times 10^{-10} $ GeV$^{-1}$ (contours at 95\% C.L.). Later analyses pointed out that the CIBER observations can instead be reasonably explained by  varying the extragalactic background light model within reasonable assumptions~\cite{Long:2019nrz,Guo:2020kiq}.
On the other hand, anomalous transparency of extragalactic sources can also represent a hint for ALP-photon conversion~\cite{DeAngelis:2008sk,Horns:2012fx}, and indeed allows to set a lower-limit in the ALPs parameter space~\cite{Horns:2012fx}.
While these two hints may partially overlap with the PSR signal region, we will not discuss them in more detail since they are partially still allowed by the CAST limit and therefore do not suffer the same tension as the PSR ALPs hint.

\subsection{Updated PSR signal region with {\it Fermi}-LAT}\label{sec:updatePSR}
In this section, we show how the PSR signal region of Ref.~\cite{Majumdar:2018sbv} is affected by uncertainties on distance 
measurements and Galactic magnetic field parameters. These parameters are included in the likelihood as additional constrained parameters, so that we can fully profile over the corresponding uncertainties.

We run a combined analysis of the gamma-ray spectra of a sample of six bright Galactic PSRs, as presented in~\cite{Majumdar:2018sbv}.
In Appendix~\ref{app:psr_sample}, Tab.~\ref{Tab:psr}, we quote the most relevant information for the PSR sample used in the present analysis.
As in~\cite{Majumdar:2018sbv}, we compare the null-hypothesis (i.e.~absence of photon-ALPs conversion) fits of the PSRs' gamma-ray spectra with the alternative hypothesis of the presence of photon-ALPs conversion in the Galactic magnetic field.
To this end, we implement the Galactic magnetic model from Jansson and Farrar~\cite{Jansson:2012pc}, with parameters updated to the latest Planck results (model Jansson12c in~\cite{Planck_GMF}).

First, we perform a combined analysis of the six objects, without including any additional constrained parameter.

Following~\cite{Majumdar:2018sbv}, we run fits of the PSRs spectra under different hypotheses, namely $H_0$ (null-hypothesis) and $H_2$ (presence of photon-ALP conversion with global parameters $m_a$ and $g_{a\gamma}$).
For this analysis, the test statistic (TS) is defined as: 
\begin{equation}
    \mathrm{TS(m_{a},g_{a\gamma})} = -2 \mathrm{ln} \bigg( \frac{\mathcal{L_{\mathrm{H_{0}}}}}{\mathcal{L}_{\mathrm{H_{2}}}} \bigg) = (\chi^{2}_{\mathrm{H_{0}}} - \chi^{2}_{\mathrm{H_{2}}}) \, .
\end{equation}
The global best-fit has TS$_{\rm max}$ = 29.08, with best-fit parameters $m_a = (4.55_{-0.13}^{+0.12})$ neV and $g_{a\gamma} = (12.0_{-1.5}^{+1.0}) \times 10^{-11}$ GeV$^{-1}$ (statistical errors only). 
The new global best fit is slightly shifted with respect to the original results~\cite{Majumdar:2018sbv}, and differences arise from updates in (i) the Jansson and Farrar magnetic field numerical implementation, and (ii) the distance estimates for the six PSRs to the latest results of the ATNF catalog\footnote{\url{https://www.atnf.csiro.au/research/pulsar/psrcat/}}. The new distances can be found Tab.~\ref{Tab:psr}.

Instead of assuming a simple $\chi^2$ distribution of the TS as in~\cite{Majumdar:2018sbv}, we here derive the distribution of the TS from the analysis (fit under $H_0$ and $H_2$) of Monte Carlo simulations of PSRs spectra under the null hypothesis. Details of the adopted statistical framework are provided in Appendix~\ref{app:chi2_psr}.
The TS distribution so obtained -- and shown in Fig.~\ref{fig:TS}, in Appendix~\ref{app:chi2_psr} -- can then be used to build the 95\% C.L.~contour region.  
Using the F-test statistic, we find the significance of the signal ($H_2$) to be about 3.5$\sigma$, so reduced with respect to the original work~\cite{Majumdar:2018sbv}.
The results are shown in Table \ref{tabl:ftest}. Additionally, we compute the TS value, $\mathrm{TS_{95\%}}$, at which one can accept the ALPs hypothesis at 95\% confidence, namely $\mathrm{TS_{95\%}} = 3.86$. 
We use this to draw the 95\% C.L.~contours of the PSR signal region.
The PSR signal region (contours at 95\% C.L.), together with the one from~\cite{Majumdar:2018sbv}, is shown in Fig.~\ref{fig:PSR_region} (left panel).  

Next, we include PSR distances as nuisance parameters in the likelihood.
To profile over the distance uncertainty for each PSR, we use the full distance probability distribution function (PDF), directly  implementing the publicly available code released by~\cite{Bartels:2018xom}.\footnote{Code available at \url{https://github.com/tedwards2412/PSRdist}.} The code calculates PDFs for distances to PSRs, and implements the latest YMW16 electron density model
~\cite{yml16}. To reduce the random noise due to Monte Carlo sampling, we run a large number of Monte Carlo simulations to reduce this effect on our PDF. 
If we consider the full width at half maximum (FWHM) as a proxy for
distance uncertainties, we can see from Tab.~\ref{Tab:psr} that FWHM is about $0.2-0.4$ kpc for the closest four PSRs (see also Fig.~\ref{fig:GMF_psr}), while it increases up to about 1 kpc for the farthest J2021+3651.
In Appendix~\ref{app:psr_sample}, Fig.~\ref{fig:psr_pdf}, we show the 
distance PDFs. We notice that, while the closest four PSRs have almost symmetric distance PDFs, the PDFs for J2240+5832 and J2021+3651 instead show long tails up to 11 and 15 kpc, respectively. 
By implementing the full PDF in our likelihood, we are able to account for all features in the distance PDF. 

With the distance uncertainty included, the $\chi^{2}$ expression, for each PSR, gets an additional term of the form: $\chi^2_{\rm dist} = -2 \, \ln \text{PDF(d)} $, 
where $\text{PDF(d)}$ is the distance PDF. 
We note that, as mentioned also in~\cite{Bartels:2018xom}, the peak of the PDF is usually very close to the ATNF estimated distance, but may not coincide with it exactly. Nevertheless, the difference is small enough that the $\chi_{\rm dist}^{2}$ does not change appreciably. 

In Fig.~\ref{fig:PSR_region} (central panel), we overlay the 95\% C.L. contours obtained when 
adding the distance of each PSR as nuisance parameter, following the same procedure highlighted above.
We can see that the best fit is now at $m_a = (4.10_{-0.17}^{+0.11})$ neV and $g_{a\gamma} = (18.5_{-2.2}^{+1.65}) \times 10^{-11}$ GeV$^{-1}$.

The best-fit distances at the global minimum are very close to actual distance estimates for all PSR but J2021+3651. In~\cite{Majumdar:2018sbv} it was already noted that J2021+3651 has the most pronounced effect on the fit due to distance uncertainty, and we observe the same happening here. The best-fit distance in this case is found to be 8.95 kpc (with a systematic uncertainty of 0.1 kpc), to be compared with the 10.51 kpc ATNF estimate.

Finally, we consider uncertainties in the Galactic magnetic field 
parameters.
We stress that searches for spectral distortions in Galactic sources
are sensitive to the product of transverse magnetic field and ALP-photon
coupling, and that there is always some degree of degeneracy between 
the best-fit ALP parameters and magnetic field ones.

Despite some new recent measurements, the Galactic magnetic field remains poorly understood and difficult to model, especially its transverse component.
For a recent review see~\cite{2019Galax...7...52J}.
We can indeed see from Tab.~1 of~\cite{Jansson:2012pc} that there is quite a large variation in some of the Galactic magnetic field parameters like $b_1, b_7, b_8$, and $z_0$. 
We here note that $b_8$ is not an independent parameter, rather it depends on all other seven spiral arms values to conserve magnetic flux, as explained in~\cite{Jansson:2012pc}.
Given the uncertainties at play, we expect the variation of at least some of these parameters to have a substantial effect on our fit. To identify what parameters affect the global fit the most, we first study the effect of varying, individually, magnetic field parameters on the single PSR best fits. Only parameters which give us maximal variation in $\chi^2$ are retained as relevant for the analysis. We find that only spiral arm field strengths $b_1, b_2, b_3, b_4$ and $b_7$ affect our best fit significantly. Effects due to other parameters are subdominant and hence we do not consider their uncertainty in what follows. 

\begin{figure}
\center
\includegraphics[scale=0.8]{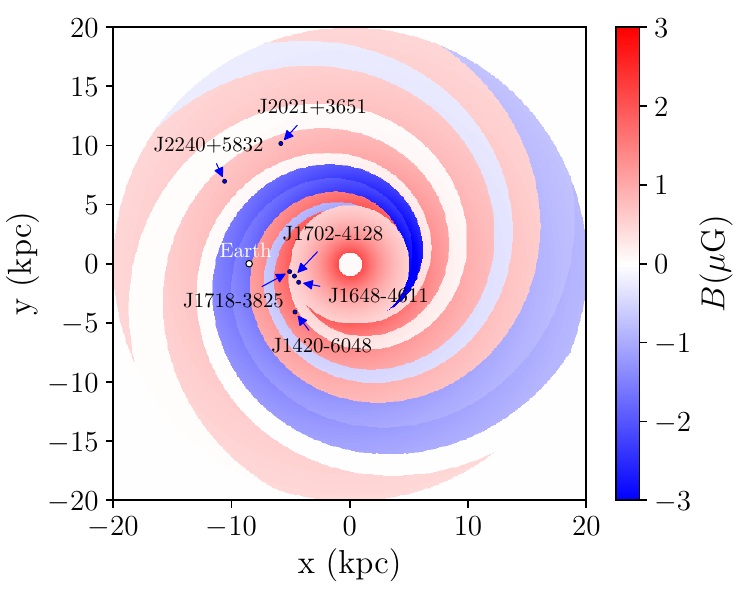}
\caption{PSR sample of interest overlaid to the Jansson and Farrar Galactic magnetic field model (Planck update), with field strength represented by the color scale.}
\label{fig:GMF_psr}
\end{figure}

We then test the dependence of the PSR signal region on the variation of 
spiral arms magnetic field parameters.
Notably, as can be seen in Fig~\ref{fig:GMF_psr}, we can distinguish two subsets of PSRs based on their position. PSRs J1718-3825, J1702-4128, J1648-4611 and J1420-6048 lie on the first spiral arm and their line-of-sight never crosses the 7$^{th}$ and 8$^{th}$ spiral arms. Hence, we expect these PSRs' fit to be unaffected by $b_7$ and $b_8$ and, indeed, we observed this in our exploratory analysis of single PSRs' best fits. Similarly, J2240+5832 and J2021+3651 lie on the 8$^{th}$ spiral arm and are unaffected by $b_1, b_2, b_3$, and $b_4$ (although those will affect the fit through $b_8$). 
Each spiral arm parameter is added to the global $\chi^2$ expression assuming a 
normal distribution with mean and variance from~\cite{Jansson:2012pc,Planck_GMF}.
Namely, the additional term writes as:
\begin{equation}
    \chi^2_{\rm arms} = \sum_{i = 1,2,3,4,7} \frac{(b_{i}-\overline{b_{i}})^{2}}{\sigma_{b_{i}}^{2}} \, ,
\end{equation}
where, $\overline{b_{i}}$ and $\sigma_{b_{i}}$ are the mean values and errors of these parameters.

In Fig.~\ref{fig:PSR_region} (right panel), we show the 95\% C.L. contours obtained when 
adding the spiral arms as nuisance parameters in the global analysis.
With these additions the new 95\% C.L region is shifted towards slightly higher couplings and lower masses, with best fit at $m_a = (4.0_{-0.10}^{+0.21})$ neV and $g_{a\gamma} = (19.75_{-2.48}^{+2.22}) \times 10^{-11}$ GeV$^{-1}$. 
The variation of spiral arms parameters at the global best-fit position is within 1$\sigma$ for all spiral arms. 

In what follows, we use as a reference PSR signal region the one obtained when profiling 
over magnetic field uncertainties, with best fit $m_a = 4$ neV and $g_{\rm PSR} = 1.97 \times 10^{-10}$ GeV$^{-1}$.

\begin{figure}
    \center
    \includegraphics[width=0.9\linewidth]{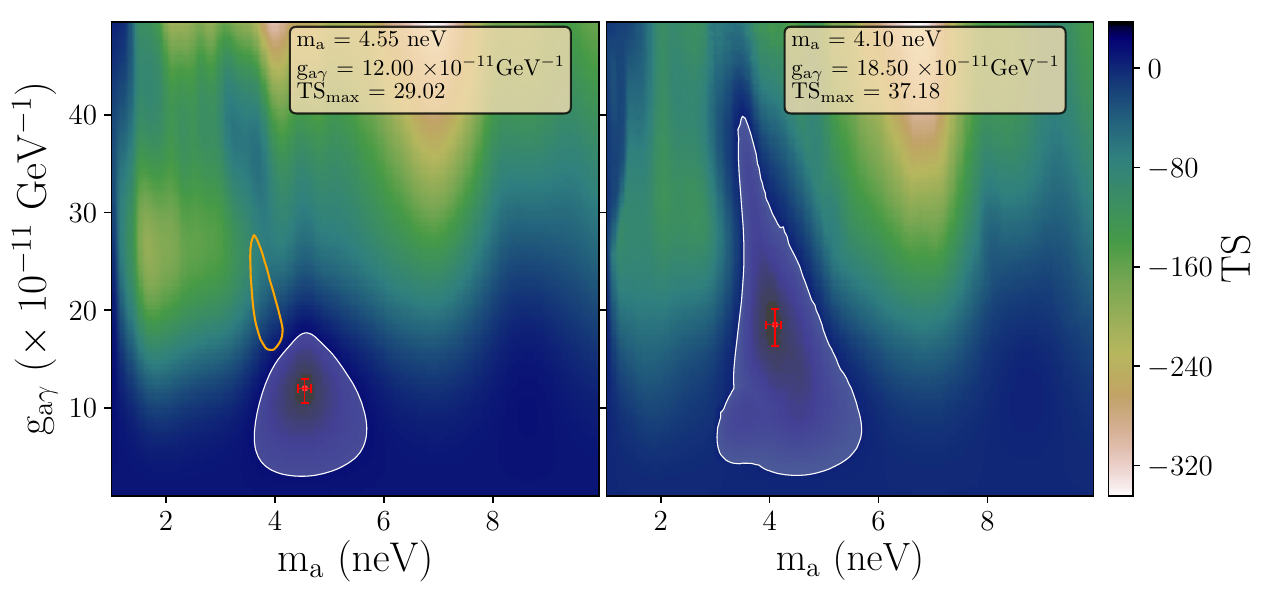} \\
    \includegraphics[width=0.55\linewidth]{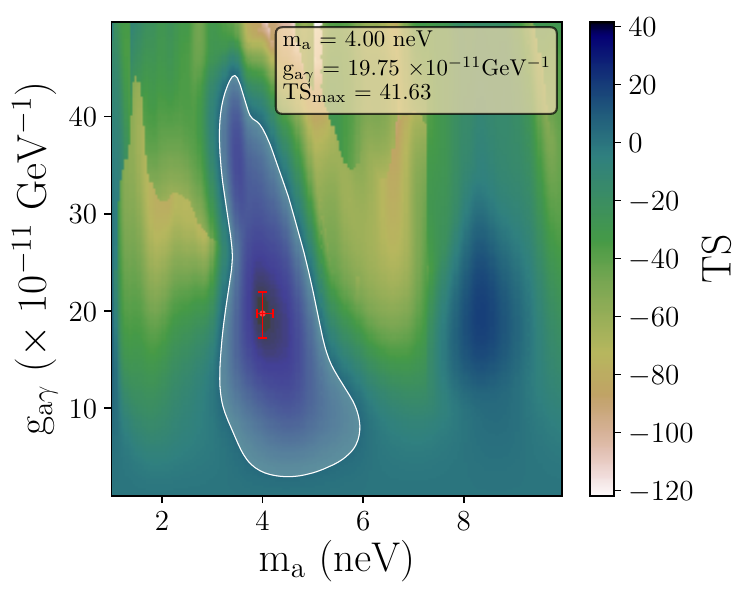}
    \caption{Updated PSR signal region, including profiling over distance and magnetic field uncertainties. \emph{Top left panel}: 95\% C.L. contours (white contours) in the $g_{a\gamma} - m_a$ plane for a global fit analysis as in~\cite{Majumdar:2018sbv}. We also overlay the original signal region from~\cite{Majumdar:2018sbv} (orange contour). We explicitly quote minimum $\chi^2$, and  $m_a$, $ g_{a\gamma}$ best-fit values, also marked by the 1$\sigma$ errors red cross.  \emph{Top right panel}: Same as left panel, when profiling over distance uncertainty of each PSR. \emph{Bottom panel}: Same as left panel, when profiling over uncertainties of the Galactic magnetic field spiral arms' parameters.}
    \label{fig:PSR_region}
\end{figure}

 %%%%%%%%%%% 
  \subsection{Bounds on ALP-photon coupling}
  \label{sec:bounds}

  %%%%%%%%%%%
  
\begin{figure}
    \centering
    \includegraphics[width=0.8\linewidth]{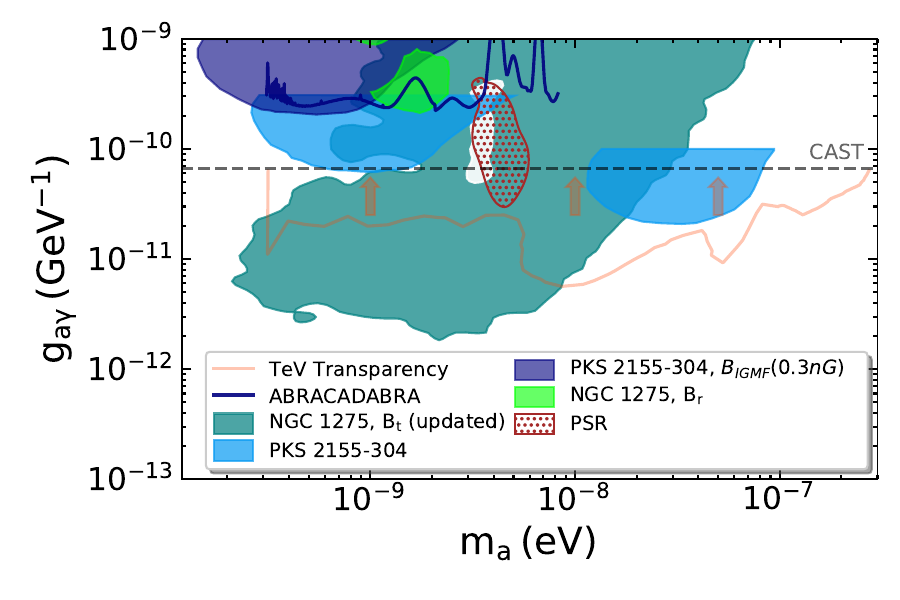}
    \caption{Comparison between the main constraints (filled regions) and the PSR signal region (hatched region, profiled over magnetic field uncertainties) from searches of high-energy oscillation features in the spectra of high-energy gamma-ray emitters. The lower limit of TeV-transparency is represented by the orange solid contour. The grey horizontal dashed line indicates the  limit derived with the CAST helioscope. We also report current limits from the ABRACADABRA laboratory experiment with the blue solid line~\cite{re: ABRACADABRA}.} 
    \label{fig:astro_constraints}
\end{figure}

A summary of the main astrophysical constraints described below, together with the updated PSR signal region, can be found in Fig.~\ref{fig:astro_constraints}.

In this section, we collect a list of references which use the spectra of GeV to TeV gamma-ray sources to set bounds on the ALP-photon coupling looking for spectral modulations and modeling the propagation in Galactic and extragalactic magnetic fields.

As for Galactic sources, as mentioned above, Ref.~\cite{Xia:2019yud} derived 95\% C.L. limits on the ALPs parameter space based on the combined analysis of three supernova remnants.
    For the same magnetic field adopted by Ref.~\cite{Majumdar:2018sbv}, the best-fit region from the PSR analysis remains still viable (red contours in Fig.~7 of \cite{Xia:2019yud}).
Ref.~\cite{Liang:2018mqm} set 95\% C.L. limits from a combined analysis of ten Galactic sources (yellow region in Fig.~2 therein). The limits pertain to masses around 100 neV, and therefore are not relevant for the PSR signal region.

Considering $> 100$ TeV (or sub-PeV) energies, Ref.~\cite{Bi:2020ths} searched for ALP-photon conversion-induced spectral modulation in the Tibet AS$\gamma$, HAWC, HEGRA and MAGIC observations of the Crab Nebula. Having found less than 1$\sigma$ improvement of the fit with ALP-photon conversion, the authors set upper limits in the ALPs parameter space: The 95\% exclusion region touches masses $m_a \sim 100 - 1000$ neV and $g_{a  \gamma} 
   \sim 10
   ^{-10} - 10^{-9} $ GeV$^{-1}$ (cf.~Fig.~3 of \cite{Bi:2020ths}), and is therefore not affecting the PSR signal hint.

Limits from extragalactic sources have been set by searching for spectral features (oscillations) in X- and gamma-ray data and required a modeling of the conversion in the intra-cluster, extragalactic and Galactic magnetic fields. X-ray constraints touch low ALPs masses (becoming very sensitive below $10^{-12}$ eV~\cite{Reynolds:2019uqt}) and therefore we do not present them here, while gamma-ray constraints are of relevance for the PSR signal region.

Two main sources have been studied in this context: NGC 1275 and  PKS 2155-304.

Using the {\it Fermi}-LAT spectrum of NGC 1275, several works set some of the strongest upper limits on ALPs from astrophysical objects~\cite{TheFermi-LAT:2016zue,Malyshev:2018rsh}. In general, the poor observational constraints on the Perseus cluster magnetic field motivated the use of intra-cluster magnetic field models built from other galaxies observations, assuming only the presence of a turbulent magnetic field component. However, \cite{Libanov:2019fzq} recently showed that including a large-scale ordered magnetic field component can alter substantially the limits, making them even much less constraining than CAST if a purely regular field is considered. Such a magnetic field component is found in several galaxy clusters and can better explain the observation of Faraday rotation measurements, while only a turbulent component fails in doing so. The model used by \cite{Libanov:2019fzq} is instead consistent with rotation measurements, X-ray observations of large-scale structures in Perseus, as well as with numerical simulations of intra-cluster magnetic fields.
In Sec.~\ref{sec:updateNGC}, we derive updated bounds from NGC 1275 using {\it Fermi}-LAT data.

Analogous limits have been set by looking at the H.E.S.S. TeV spectrum of PKS 2155-304 \cite{Abramowski:2013oea}. Ref.~\cite{Zhang:2018wpc} used {\it Fermi}-LAT observations of PKS 2155-304 to set bounds on ALPs which overlap with the GeV constraints from NGC 1275. Again, only a purely turbulent magnetic field in the cluster is used in the two works. Ref.~\cite{Bu:2019qqg} tested how much the limits from the {\it Fermi}-LAT spectrum of PKS 2155-304 are impacted by different choices of magnetic field models. 
Finally, \cite{Guo:2020kiq} set constraints from TeV spectra of PKS 2155-304 and PG 1553+113 (Fig.~6 in \cite{Guo:2020kiq}), which again partially overlap with {\it Fermi}-LAT bounds from NCG 1275.
In Sec.~\ref{sec:updatePKS}, we derive updated bounds from PKS 2155-304 using H.E.S.S.~data.

\subsection{Updated bounds from NGC~1275 with {\it Fermi}-LAT}\label{sec:updateNGC}
The published bound from \cite{TheFermi-LAT:2016zue}, which derived limits using {\it Fermi}-LAT observations of NGC 1275,  was calculated for a two-photon coupling $g_{a\gamma}$ smaller than
the CAST bound. In order to update this bound with more recent data and analysis/calibration, we have carried out a dedicated
analysis using the same approach as suggested in \cite{TheFermi-LAT:2016zue},  but extending the limit calculation to larger coupling constants 
in order to cover the region favored by the analysis of Galactic PSR spectra. 
In particular, we extend the parameter space to the previously unexplored range of coupling  $g_{a\gamma}>7\times10^{-11}~\mathrm{GeV}^{-1}$. 

The data-set used here includes photons recorded between 2008-08-04 until 2020-09-09 with maximum zenith angle of $90^\circ$ in the energy range between 100 MeV and 500 GeV with 30 logarithmically spaced bins per energy decade.
We use \texttt{fermipy} 1.0.0., interfaced with Fermitools~\texttt{P8R3\_V3}.
%Software versions are Fermitools~\texttt{P8R3\_V3} and \texttt{fermipy} 1.0.0.
We select a region of interest of $10^\circ\times 10^\circ$ centered on NGC~1275
and extend the model up to a radius of $15^\circ$. Spatial bin size is $0.2^\circ$. Fitting of the spectra without ALP-contribution is done by repetitive calls of the \texttt{fermipy} method \texttt{optimize}, followed by \texttt{fit}, where fitted sources are NGC 1275, Galactic diffuse emission, isotropic diffuse emission and normalization of all sources within $3^\circ$ or $TS > 100$. The loglike of the null hypothesis $\mathcal{L}_0$ is taken after these fitting steps by the \texttt{fermipy} method \texttt{like}.
We adopt a purely turbulent intra-cluster magnetic field as in~\cite{TheFermi-LAT:2016zue}.
%Added the spatial and spectral bin sizes, max zenith angle - Julian

For each pair of $m_a$, $g_{a\gamma}$
from a logarithmically  spaced $30\times 30$ square grid in the range
of $m_a\in [10^{-10},10^{-7}]~\mathrm{eV}$ and $g_{a\gamma} \in [10^{-12},10^{-9}]~\mathrm{GeV}^{-1}$ we randomly 
sample 100 realizations  $\mathbf{B}(r)$ of the  intra-cluster magnetic field  
with a turbulent power spectral density following a power law. 

For each (random) magnetic field realization, the conversion probability $P_{\gamma\rightarrow a}(E_\gamma)$ 
is calculated \cite{Meyer:2014epa} and multiplied with a log-parabola function 

\begin{equation}
    F(E_\gamma) = (1-P_{\gamma\rightarrow a}(E_\gamma))  \Phi_0 \left(\frac{E_\gamma}{E_0}\right)^{-\alpha-\beta\log(E_\gamma/E_0)}
\end{equation}
with
free parameters $\mathbf{\theta}=\{\Phi_0, \alpha, \beta\}$ and $E_0$ fixed. 
%The  likelihood for each choice of the realization of the magnetic field $\mathcal{L}(\mathbf{\theta},B,g_{a\gamma},m_a)$ is calculated  using \texttt{fermipy} \new{(starting from the above described model without ALP contribution)} and the parameters $\mathbf{\theta}$ are varied using \texttt{iminuit} to maximize the likelihood. 
%The resulting likelihood values are stored and the $95$-percentile value $\mathcal{L}_{max,95}(g_{a\gamma},m_a)$ is determined for each grid point. 

For each grid point, we determine the maximum likelihood by minimizing \texttt{fermipy}'s negative logarithmic model likelihood. Our own optimization code instructs \texttt{iminuit} to optimize said likelihood by varying $\mathbf{\theta}$, starting from the model without ALPs. Finally, we take the $95$-percentile value $\mathcal{L}_{\rm max,95}(g_{a\gamma},m_a)$ of the 100 sampled magnetic fields for use in the TS distribution.

\begin{figure}
    \centering
    \includegraphics[width=0.7\linewidth]{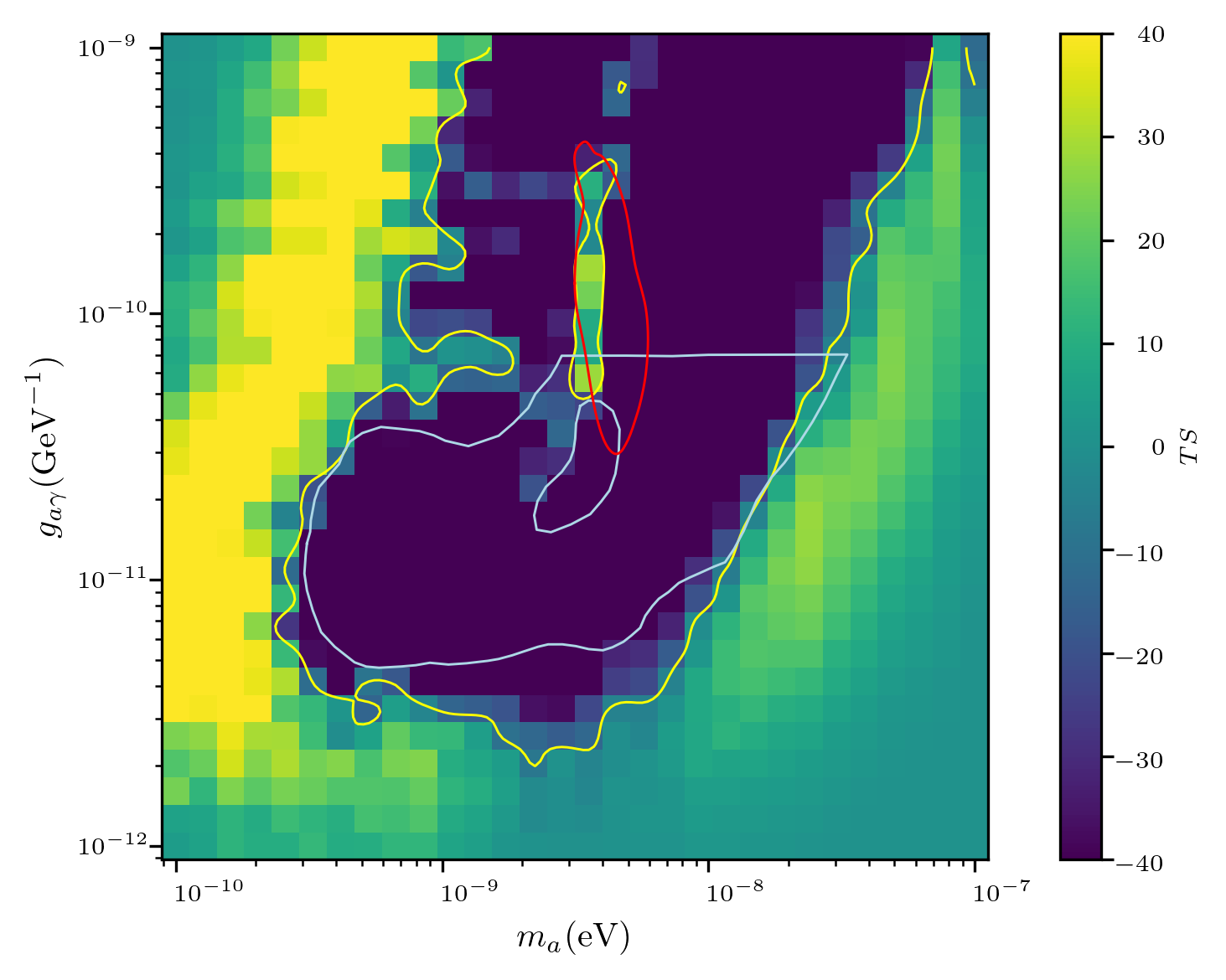}
    \caption{Updated ALPs bounds from the {\it Fermi}-LAT spectrum of NGC 1275. The scan over the parameter space has been extended to couplings higher than the CAST limits so to test the validity of the PSR ALPs signal hint (red contour, profiled over magnetic field uncertainties). The colorbar indicates the test statistics $TS$ (see the text for more details). A purely turbulent magnetic field has been used, to fully compare with results from \cite{TheFermi-LAT:2016zue} (light blue contour). 
    \label{fig:fermi_update}}
\end{figure}

%For the scenario with vanishing coupling, we assume $P_{\gamma\rightarrow a}=0$ and determine
%$\mathcal{L}_0(g_{a\gamma}=0)$. 
The logarithm of the likelihood-ratio
is used to define the test statistics
\begin{equation}
TS(g_{a\gamma},m_a)=2(\ln(\mathcal{L}_{\rm max,95}(g_{a\gamma},m_a)) - \ln(\mathcal{L}_{0})) \,\ 
\end{equation}
which is shown in Fig.~\ref{fig:fermi_update}
for values of $-40<TS<40$.
Since the distribution of $TS$ under the null-hypothesis is not known,
we choose to simulate mock data sets by re-sampling the expected counts in 
the data cube following the Poissonian distribution (using the \texttt{simulate\_roi} feature of \texttt{fermipy}). Each of the 100 mock data sets is then analysed in the same fashion as the real-data set. The maximum
$TS$ for each of the data sets is collected in a distribution. This procedure
follows closely the approach described in \cite{TheFermi-LAT:2016zue} with
the difference that the mock data-sets there were generated using the
full simulation of the instrumental response via \texttt{gtobssim}. The
resulting probability density distribution of the $TS$ values under 
null-hypothesis is then fit with non-central $\chi$-squared distribution with
11.25 degrees of freedom and a non-centrality of $1.1\times 10^{-5}$ for 
re-scaling factor of $0.39$.

We find a maximum value of $TS_{\rm max}=64>TS_{3\sigma}\approx 12.03$
for $(m_a,g_{a\gamma})=(4.7\times 10^{-10}~\mathrm{eV}, 2.7\times 10^{-10}~\mathrm{GeV}^{-1})$. The threshold value $TS_{3\sigma}$ has
been determined from the fit to the TS distribution from simulated
data sets. The large value of $TS$ is related to a feature at 
the low energy part of the energy spectrum. We find that this
feature is compatible with the systematic uncertainties that we
estimate from choosing different settings for the treatment of the
energy dispersion and well within 
the systematic uncertainty of the instrumental response function\footnote{\url{https://fermi.gsfc.nasa.gov/ssc/data/analysis/scitools/Aeff_Systematics.html}}. 

We nevertheless proceed with the calculation of an exclusion region
of $m_a, g_{a\gamma}$. Following the same approach as in 
\cite{TheFermi-LAT:2016zue}, we determine the threshold value for
the confidence region using the same simulation as explained above. 
The resulting contour for a confidence level of $95~\%$ is shown 
in Fig.~\ref{fig:fermi_update}. 

The exclusion region found here overlaps in a consistent way with the exclusion region found 
by \cite{TheFermi-LAT:2016zue} as indicated in Fig.~\ref{fig:fermi_update}. There are some differences at the high- and low-mass edges which 
are related to differences in the data analyses procedure and the deeper exposure of the data-set used here. 
In the low-coupling regime, the 
exclusion region found here extends by about 0.1 dex to smaller values which can be 
explained by the increased statistics of the observational data (increasing in exposure from 
6 years to 11 years). The region of fast oscillation in the spectrum at the center of the 
exclusion region found by \cite{TheFermi-LAT:2016zue}  (around the hole) is almost entirely excluded, 
which is again plausibly explained by the improved statistics. There are notable regions at low mass which 
fit better the data. This requires further investigation but it is very likely  related to a 
notable feature at the
low-energy end of the measured spectrum where uncertainties of the instrumental response function
are the strongest. 

 The signal from the modulations in energy spectra from Galactic PSR
is marked by the orange contour. The values of $TS$ found in that region are predominantly positive,  however this does not represent a significant improvement of the fit. 
Given the observations and analysis carried out here, the PSR signal region can therefore not be excluded by NGC 1275 {\it Fermi}-LAT data, even 
under the  choice of a purely turbulent intra-cluster magnetic field. 

\subsection{Updated bounds from PKS 2155-304 with H.E.S.S.}\label{sec:updatePKS}
The bound derived in \cite{Abramowski:2013oea} relates to two possible scenarios for
the magnetic field: Turbulent magnetic field in the inter-galactic medium and turbulent
magnetic field in the intra-cluster medium. In Fig.~\ref{fig:astro_constraints}, the 
resulting limits from \cite{Abramowski:2013oea} are shown for $B_{RMS}=1~\mu$G in the cluster, 
and $B_{RMS}=1~n\mathrm{G}$ for the inter-galactic magnetic field. The latter scenario 
leads to a bound which partially excludes the signal region favored by the PSR data set. However, that
particular scenarios 
is  quite optimistic given that a recent analysis of the anisotropy of the cosmic-microwave background
constrains the primordial magnetic field to be $<0.047$~nG \cite{Jedamzik:2018itu}. We include in Fig.~\ref{fig:astro_constraints} 
the resulting constraint for $B_{RMS}=0.3$~nG to demonstrate that the bound is considerably relaxed for a more realistic
choice of the inter-galactic magnetic field.

%%%%%%%%%%%%%%%%%%%%%%
\section{Dynamical suppression of the solar ALP flux}
  \label{sec:dynamical_suppression}

%%%%%%%%%%%%%%%%%%%%%%%%

\subsection{Standard solar ALP flux}
\label{sec:solar_ALP_flux}

The ALP-two photon vertex is described by the Lagrangian term
%....................................................................
\begin{equation}
{\cal L}_{a\gamma}=-\frac{1}{4} \,g_{a\gamma}
F_{\mu\nu}\tilde{F}^{\mu\nu}a=g_{a\gamma} \, {\bf E}\cdot{\bf B}\,a~,
\label{eq:lagrangian}
\end{equation}
%.....................................................................................
where $g_{a\gamma}$ is the ALP-photon coupling constant (which has dimension of an inverse energy),
$F$ the electromagnetic field and $\tilde F$ its dual.

The primary production mechanism for ALPs interacting with photons in the core of the Sun is
the Primakoff process $\gamma + Ze \to Ze +a$, where a thermal photon at a temperature $T$ in the stellar core 
converts into an axion in the Coulomb fields of nuclei and electrons.
The transition rate for a photon of energy $E$ into an ALP of the same energy by the Primakoff effect in a stellar plasma is~\cite{Raffelt:1987yu}
%....................
\begin{equation}
\Gamma_{\gamma\to a}= \frac{g_{a\gamma}^2 T \kappa_s^2}{32 \pi}\left[\bigg(1+\frac{\kappa_s^2}{4 E^2} \bigg)
\ln\bigg(1+ \frac{4E^2}{\kappa_s^2} \bigg)-1\right] \,\ ,
\label{eq:primakoff}
\end{equation}
%........................
where $T$ is the plasma temperature.
Recoil effects are neglected so that the photon and axion energies are taken to be equal.
The function $\kappa_s$ is the Debye-H{\"u}ckel screening scale 
 %.......................
 \begin{equation}
 \kappa_s^2 = \frac{4 \pi \alpha}{T}\frac{\rho}{m_u}\left(Y_e+ \sum_j Z_j^2 Y_j \right) \,\ ,
 \label{eq:screen}
 \end{equation}
 %.....................
with $\rho$ the mass density, $m_u=1.66\times 10^{-24}\,{\rm g}$ the atomic mass unit, 
$Y_e$ the number of electrons per baryon, 
and $Y_j$ the number (per baryon) of 
the ions with nuclear charge $Z_j$.
At low density, the electrons are non-degenerate and the ion correlation can be neglected, so that 
the Debye-H{\"u}ckel theory provides a valid description of the plasma screening. 
This condition is certainly fulfilled in the solar core.
One should also take into account the plasma frequency for the photons in the system, $\omega_{\rm pl}^2\simeq 4 \pi \alpha n_e/m_e$,
entering the photon  dispersion relation $k=\sqrt{E^2-\omega_{\rm pl}^2}$ where
$k$ is the photon momentum. The value of the plasma frequency depends on the radial position in the Sun.

Integration over the whole Sun gives the number of emitted ALPs per unit time~\cite{Andriamonje:2007ew}
%.......................
\begin{equation}
\Phi_a^0=R_{\odot}^3 \cdot \int_{0}^{1} dr \, 4 \pi r^2 \int_{\omega_{\rm pl}}^{\infty} dE \, \frac{4 \pi k^2}{(2\pi)^3} \frac{dk}{dE} \, 2 f_B \Gamma_{\gamma \to a} \,\ ,
\label{eq:ax_spectrum}
\end{equation}
%...................
where $f_B=(e^{E/T}-1)^{-1}$ is the Bose-Einstein distribution of the thermal photon bath in the solar plasma and 
$r=R/R_{\odot}$ is  a dimensionless solar radial variable, normalized to the solar radius $R_{\odot}=6.9598\times 10^{10}$~cm.

%%%%%%%%%%
\begin{figure}[t!]
\center
\includegraphics[width=0.45\textwidth]{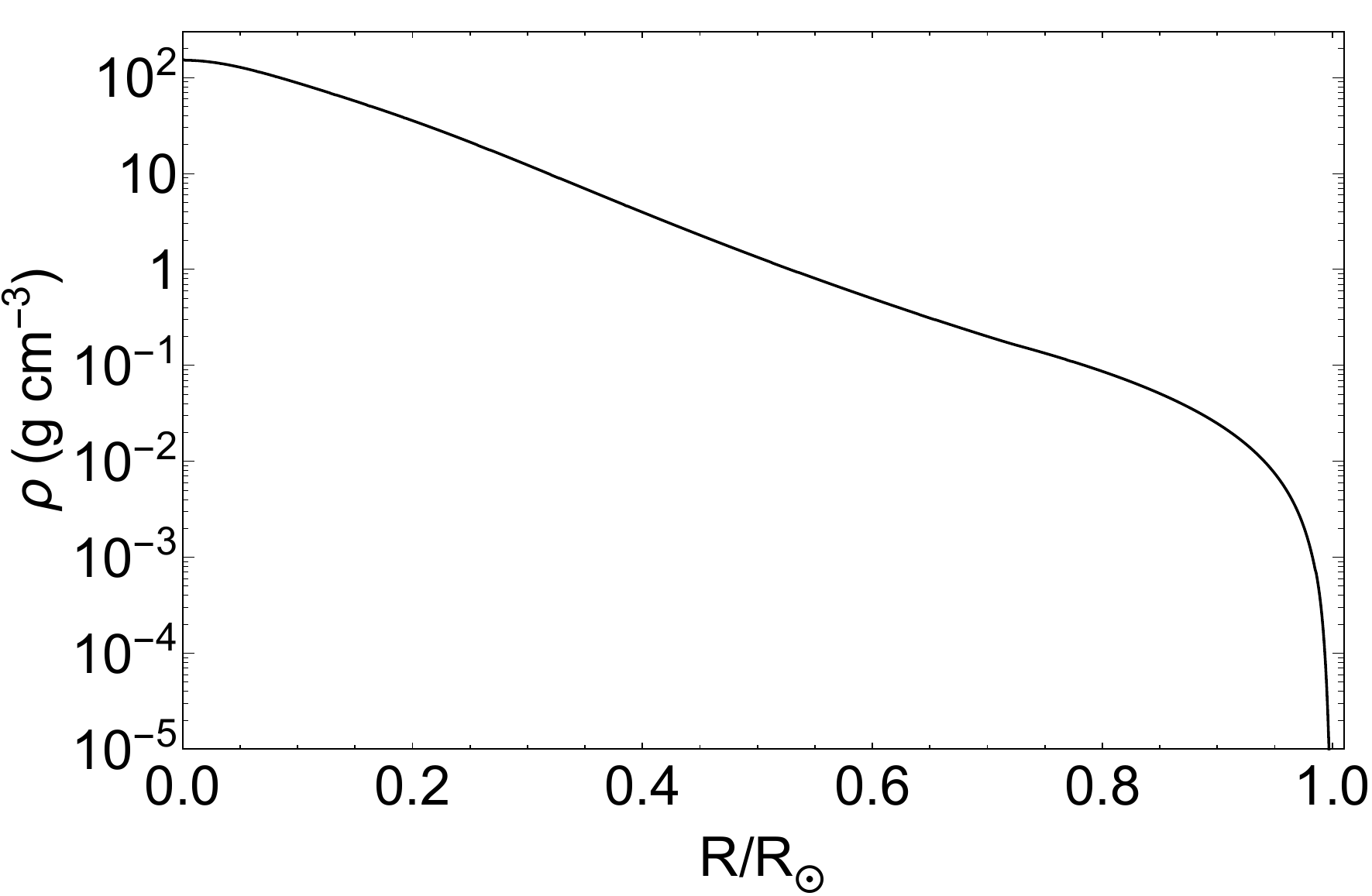}
\includegraphics[width=0.45\textwidth]{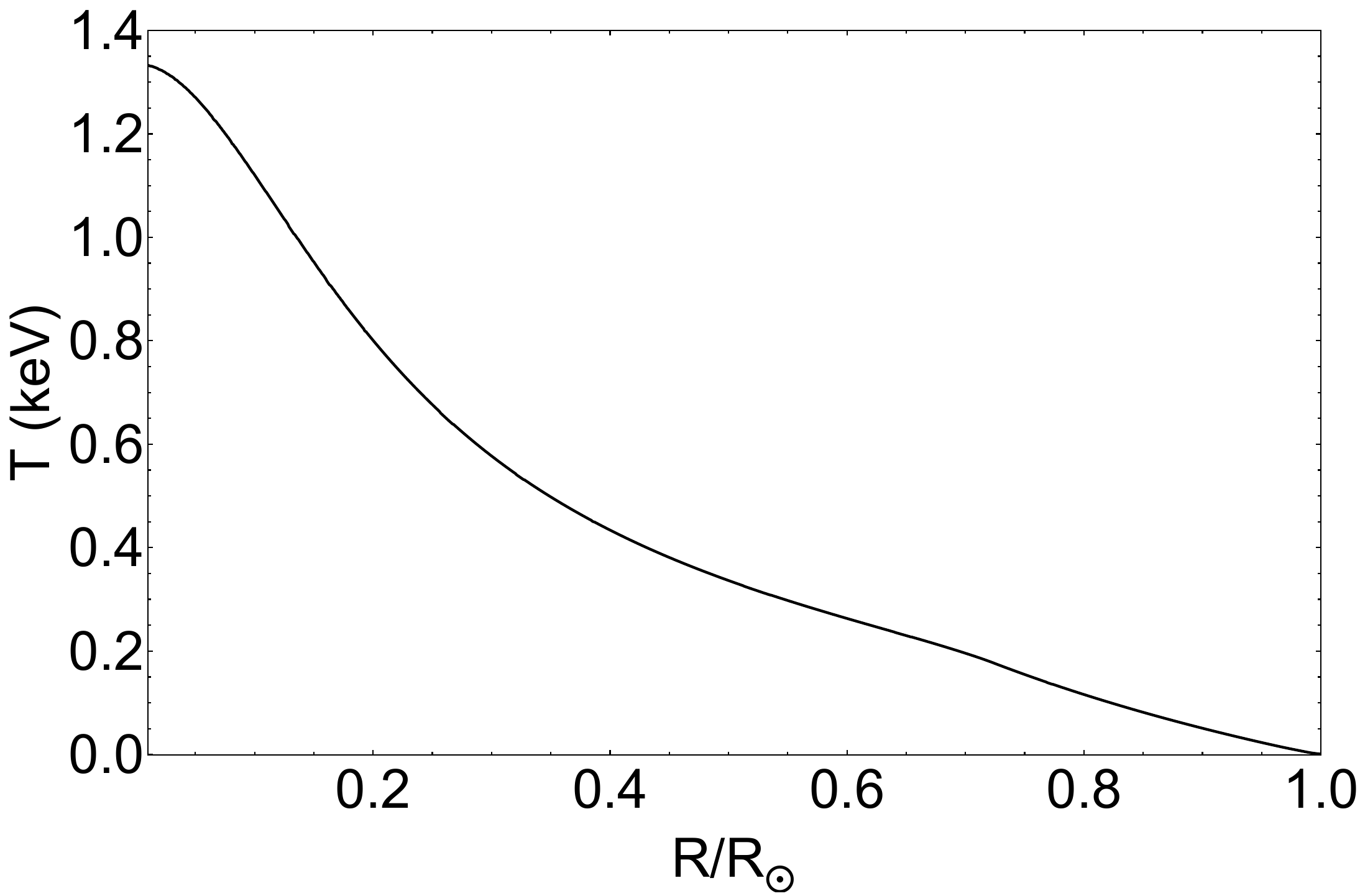}
\caption{Standard solar model AGSS09. Radial profiles of density $\rho$ ({\it left panel}) and temperature $T$ ({\it right panel}).
}
\label{fig:sun_model}
\end{figure}
%%%%%%%%%%%%

\begin{figure}[t!]
\center
\includegraphics[width=0.6\textwidth]{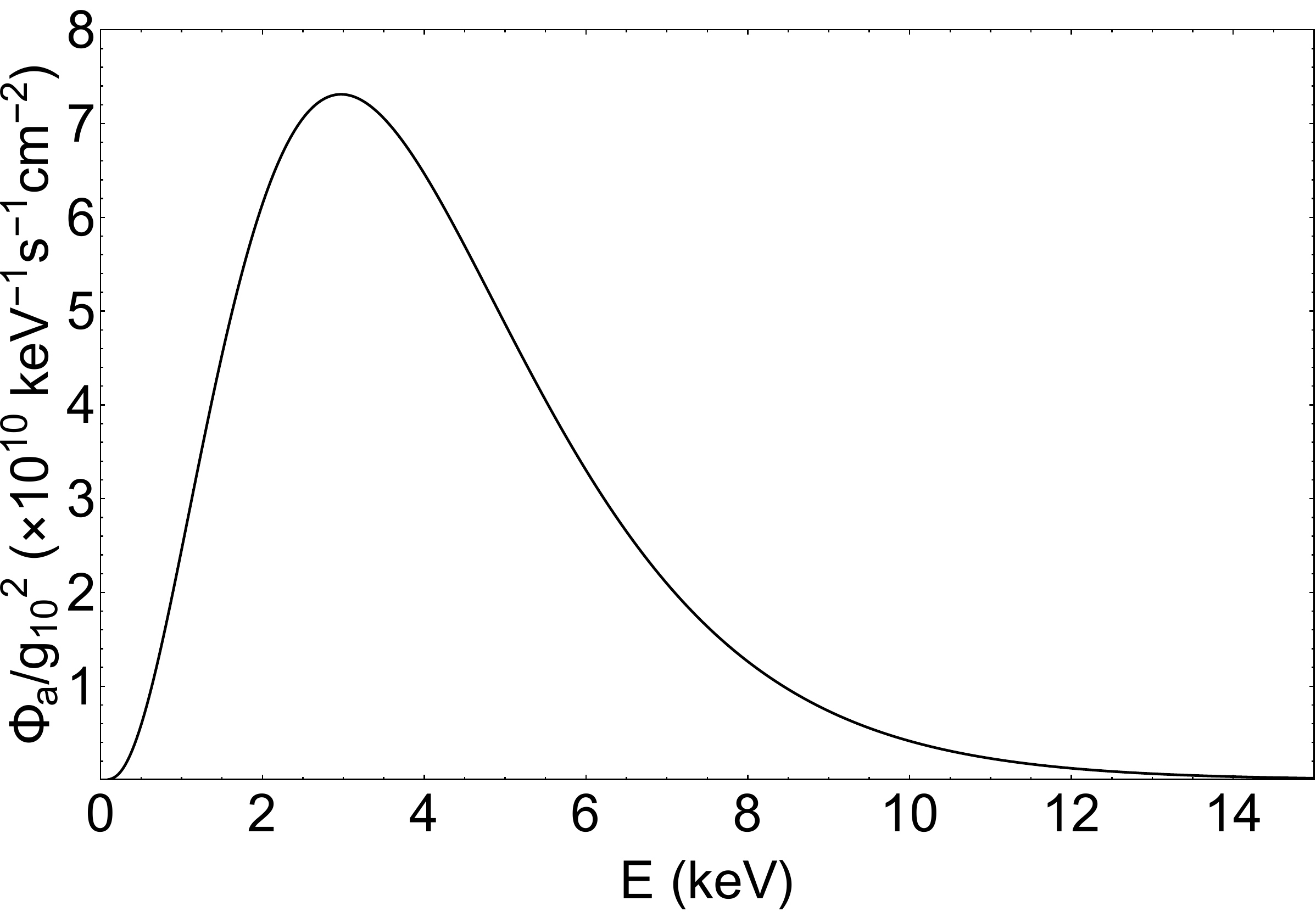}
\caption{Standard solar ALP flux at Earth from the Primakoff process, normalized to the value of $g_{10}\equiv g_{a\gamma}/10^{-10}$~GeV$^{-1}=1$.}
\label{fig:alp_fluxs}
\end{figure}

In Fig.~\ref{fig:sun_model}  we show the radial profiles of  the solar matter density $\rho$ (left panel) and temperature  $T$ (right panel) from the 
standard solar model AGSS09~\cite{Serenelli:2009yc}. Integrating over these radial profiles from Eq.~(\ref{eq:ax_spectrum})  one can calculate the solar ALP spectrum, 
 shown in Fig.~\ref{fig:alp_fluxs}. This is in agreement with the 
usual one published in literature (see, e.g.,~\cite{Andriamonje:2007ew}).

%%%%%%%%%%
%%%%%%%%%%%%
 Following~\cite{Andriamonje:2007ew}, we used as spectral fit for the solar ALP spectrum the following 
 function
 %..................................
 \begin{eqnarray}
 \frac{d \Phi_a^0}{d E} &=& g_{10}^2 \cdot \ 10^{10} \,\ \textrm{keV}^{-1} \,\ \textrm{cm}^{-2} \,\ \textrm{s}^{-1} \,\ \nonumber \\
 &\times & C \left(\frac{E}{E_0} \right)^\beta \exp\left[-(\beta+1)E/E_0 \right] \,\ ,
 \end{eqnarray}
 %.....................................
  where $g_{10}\equiv g_{a\gamma}/10^{-10}$~GeV$^{-1}$, $C$ is a normalization constant, $\beta$ a parameter that controls the shape of the spectrum, and $E_0$ the average energy, $E_0=\langle E \rangle$.    
  Numerical values of these parameters
  are given in Table~\ref{tab:parameters}. The accuracy of the fit is better than $1\%$ in the energy window $E\in[1;10]$~keV.
  
  %%%%%%%%%%%%%%%%%%%%%%%%%%%%%%%
 \begin{table}[!t]
 \caption{Parameters of the solar ALP spectrum for different values of $g_{10}$. We used $r_c=0.50 R_{\odot}$.}
\begin{center}
\begin{tabular}{lclclc|c}
\hline
$g_{10}$ & $C$ & $E_0$ (keV) & $\beta$  \\
\hline
\hline
1 & 202.80 & 4.16 & 2.48  \\
$\theta(r-r_c)$ & 32.61 & 1.15 & 2.83 \\
\hline
\end{tabular}
\label{tab:parameters}
\end{center}
\end{table}
%%%%%%%%%%%%%%%%%%%%%%%%%%%%%%%%

%%%%%%%%%%%%%%%%%%%%%%
\subsection{Environmental suppression of the solar ALP flux}  
\label{sec:suppression_solar_flux}
 %%%%%%%%%%%%%%%%%
 
 A way to suppress the solar ALP flux and evade the CAST bound is to assume that
 the coupling constant $g_{a\gamma}$ is no longer a constant but an environment dependent
 quantity~\cite{Jaeckel:2006xm} 
 %...............
 \begin{eqnarray}
 g_{a\gamma} &\to& g_{a\gamma}(\eta) \nonumber \\ 
   \eta &=& \omega_{pl}, T, \kappa_s^2, \rho, q^2, \ldots
  \end{eqnarray} 
  where $\eta$ is an environmental parameter that might depend on plasma density, on temperature, etc. 
  
  %%%%%%%%%%
\begin{figure}[t!]
\center
\includegraphics[width=0.6\textwidth]{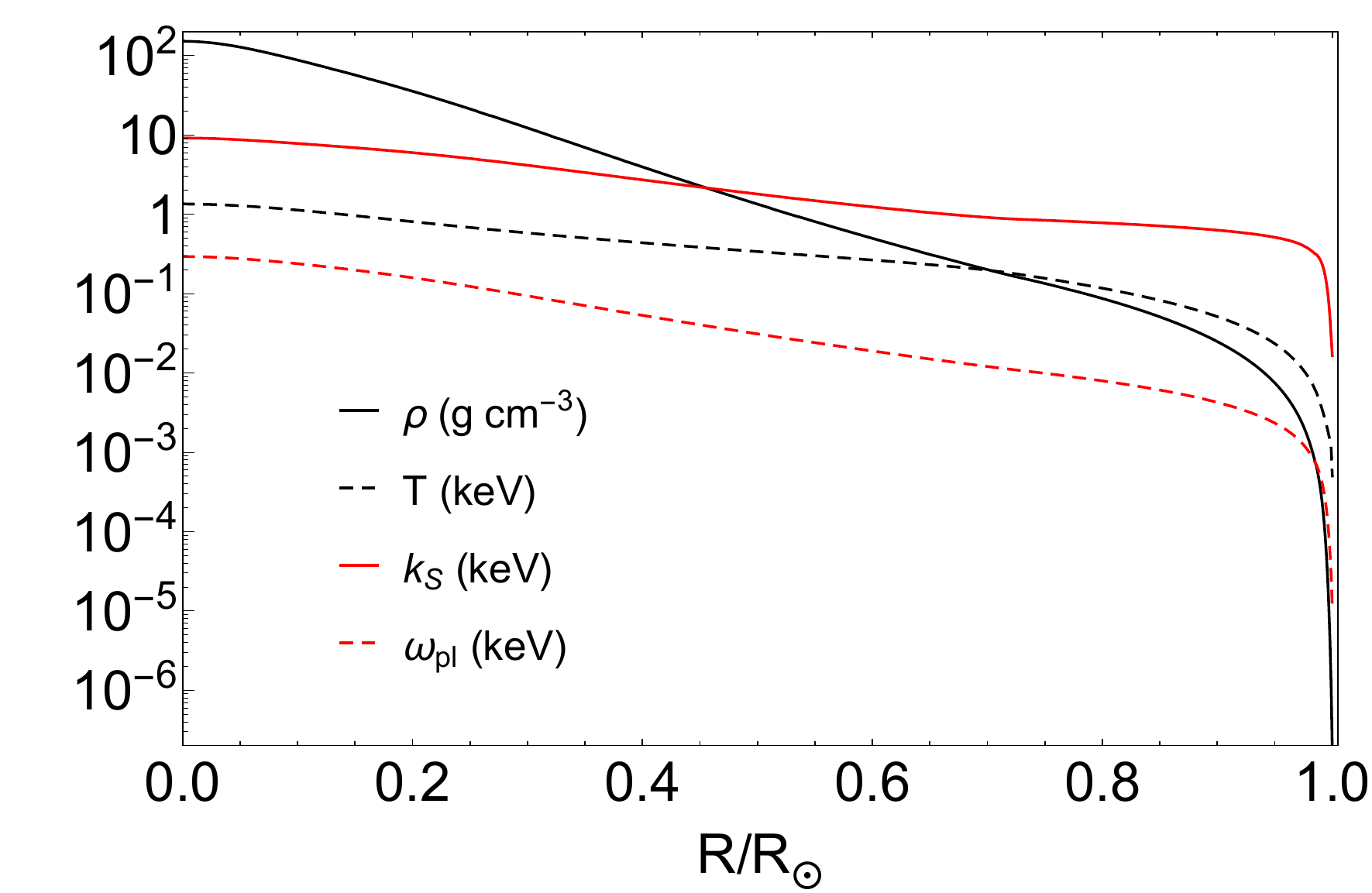}
\caption{Radial evolution of the environmental parameters in the Sun. }
\label{fig:quantities}
\end{figure}
%%%%%%%%%%%%
  
   As shown in Fig.~\ref{fig:quantities}, all solar environmental parameters 
   decrease monotonically with the radial distance from the center of the Sun. 
   Therefore, in order to reconcile the solar ALP
   bound with the PSR claim at lower densities, one should assume that $g_{a\gamma}$ is suppressed
   for high values of $\eta$.
   In particular, lacking a detailed model providing the coupling suppression, we empirically assume
   %...........................
   \begin{equation}
   g_{a\gamma}(r,r_c)= g_{\rm PSR}\,\theta(r-r_c) \,\ ,
   \label{eq:gag_rc}
   \end{equation}
   %............................
   where $g_{\rm PSR}$ is the value given by the PSR claim and $ \theta(x) $ 
   is the Heaviside step function, equal to 0 for $ x<0 $ and to 1 for $ x\geq 0 $. 
   Of course, in a realistic model one would not expect the coupling to drop sharply to zero at a certain radius.
   Nevertheless, this serves as a model independent framework to discuss our mechanism. 

Evidently, since the ALP coupling is suppressed in the core of the star, the ALP flux is going to be reduced in this scenario.
We define the flux suppression factor as~\cite{Jaeckel:2006xm}  
%......................
\begin{equation}
S(E;r_c)= \frac{d\Phi_a(E,r_c)}{dE}/\frac{d\Phi_a^0(E)}{dE} \,\ ,
\end{equation}
%.......................
where $ \Phi_a^0(E) $ is defined in Eqs.~\eqref{eq:ax_spectrum} and $ \Phi_a(E,r_c) $ has an analogous
definition but with $ g_{a\gamma}\to g_{a\gamma}(r,r_c) $, defined in Eq.~\eqref{eq:gag_rc}.

   If the flux of ALPs from a stellar plasma is suppressed by a factor $S$, in order to have a consistent scenario 
   between the PSR claim and the CAST bound one should require
   %..................
   \begin{equation}
   (S g_{\rm PSR}^2) \, g_{\rm PSR}^2 < g_{\rm CAST}^2\, g_{\rm CAST}^2 \,\ ,
   \end{equation}
   %....................
   where  the second $g_{a\gamma}^2$ factor comes from   the reconversion at Earth resulting in a total counting rate scaling as $g_{a\gamma}^4$.
   Numerically one gets
   %.........................
   \begin{equation}
   S < 1.3 \times 10^{-2} \,\ .
   \label{eq:supprval}
   \end{equation}
   %.....................

\begin{figure}[t!]
\center
\includegraphics[width=0.6\textwidth]{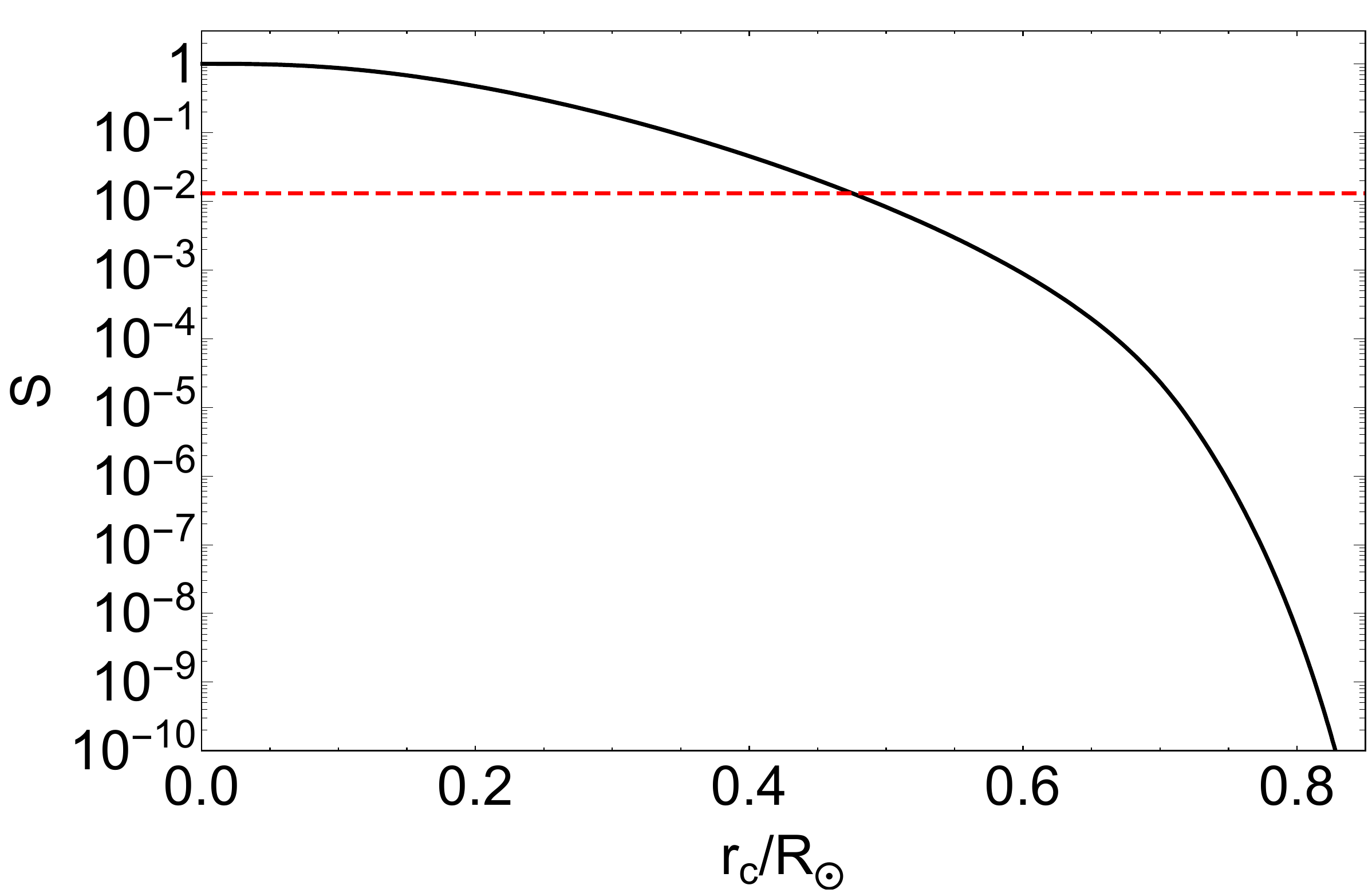}
\caption{Suppression factor of the flux of ALPs, $S\,(E=2 \,\ \textrm{keV},\,r_c)$, as a function of $r_c$.
The horizontal dashed line represents the threshold required to reconcile CAST bound with the PSR claim.}
\label{fig:suppr}
\end{figure}

   Following Ref.~\cite{Jaeckel:2006xm}, in order to maximize the suppression factor we calculate it at a fixed energy $E=2$~keV, corresponding to the lower threshold
   of the CAST analysis for solar ALPs~\cite{Anastassopoulos:2017ftl}. 
   In Fig.~\ref{fig:suppr} we show the flux suppression factor $S\, (E=2\ \textrm{keV},\, r_c)$ as a function of $r_c$.
   Evidently, in order to get the suppression value of Eq.~(\ref{eq:supprval}), one should have
   $r_c=0.50\, R_{\odot}$, corresponding to $\rho_c= 1.3$~g/cm$^{-3}$ or $T=0.34$~keV.
   
  %%%%%%%%%%
%%%%%%%%%%%%

%%%%%%%%%%
\begin{figure}[t!]
\center
\includegraphics[width=0.6\textwidth]{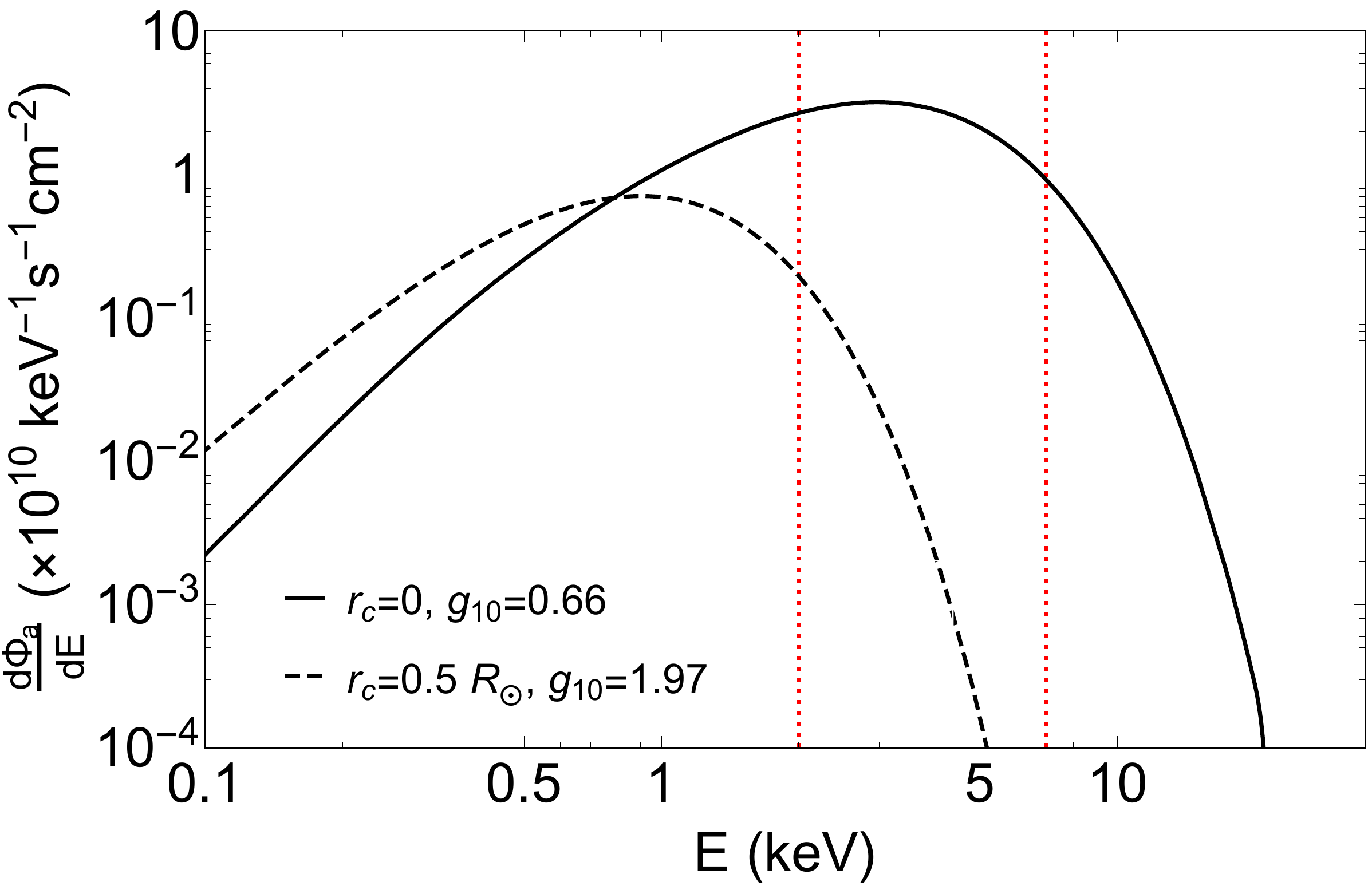}
\caption{Solar ALP flux from Primakoff process for $r_c=0$ and $g_{10}=0.66$ (continuous curve) and 
for $r_c= 0.50 R_{\odot}$ for $g_{10}=1.97$. The dashed vertical lines indicate the detection band of CAST $E\in[2;7]$~keV.}
\label{fig:alp_flux}
\end{figure}
%%%%%%%%%%%%

The resulting ALP flux, obtained integrating over the solar model for $r>r_c$ and $g_{10}=1.97$ is shown in Fig.~\ref{fig:alp_flux}    
  (dashed curve) in comparison with the standard case for $r_c=0$ and $g_{10}=0.66$ (continuous case). 
  The detection band of CAST, $E\in[2;7]$~keV, is shown as vertical dashed lines. It results that in this energy window, the ALP flux with an environmental
  dependent coupling is strongly suppressed with respect to the standard case.  
The spectral parameters in this case are shown in Table~\ref{tab:parameters}.
  
 %%%%%%%%%%%%%%%%%%%%
 \section{Phenomenological consequences}
 \label{sec:phenomenology}
 %%%%%%%%%%%%%%%%%%%% 

 %%%%%%%%%%%%%%
 \subsection{Solar ALPs at CAST}
 \label{sec:CAST}
  %%%%%%%%%%%%% 
 
  The suppression of the solar ALP flux due to environmental effects shown in Fig.~\ref{fig:alp_flux} allows one to significantly relax the 
  CAST bound from non-observation of solar ALPs. Indeed, the measurement in the energy window $E\in[2;7]$~keV
implies a  constrain $g_{10} < 0.66$ in the standard case. In order to extrapolate this bound to the case with a suppressed
  $g_{a\gamma}$, we evaluate the total new ALP flux in the energy window $E\in[2;7]$~keV and we impose
  %........................
  \begin{equation}
  \int\limits_{2\,{\rm keV}}^{7\,{\rm keV}} dE \frac{d\Phi_a}{dE}(g_{10}, r_c) <  \int\limits_{2\,{\rm keV}}^{7\,{\rm keV}} dE \frac{d\Phi_a^0}{dE}(g_{10}=0.66) \,\ ,
  \end{equation}
  %........................
  {with  $r_c=0.50 R_{\odot}$} obtaining
  %.........................
  \begin{equation}
  g_{10} < 23.7 \,\ .
  \end{equation}
  %........................
  It is evident that this significant relaxation of the CAST bound would make it compatible with the PSR ALP claim. 
  
\begin{figure}[t!]
\center
\includegraphics[width=0.6\textwidth]{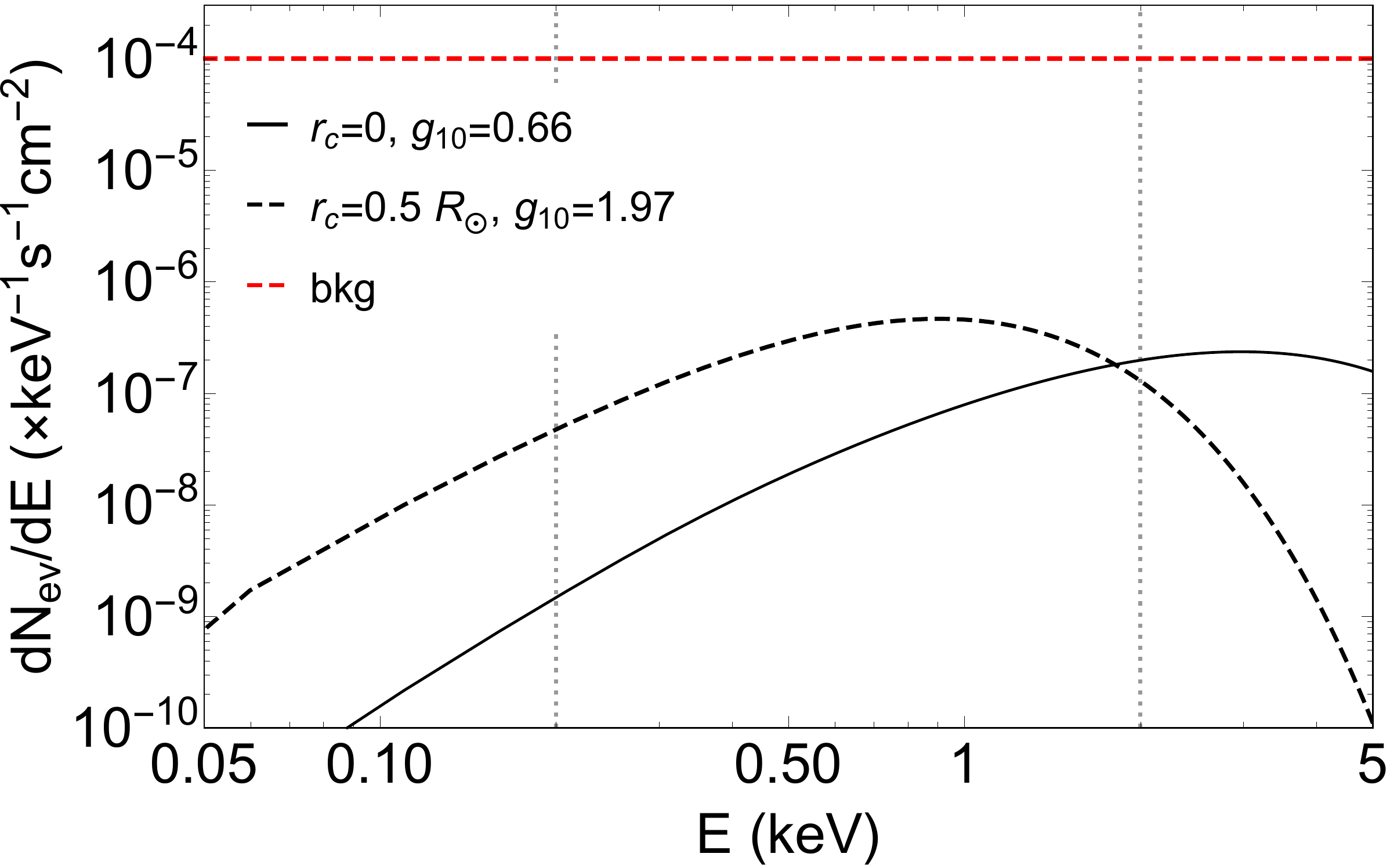}
\caption{Event rate in CAST for low-energy solar ALP flux in the energy-window $E\in[0.2:2]$~keV for the standard case (continuous curve) 
and the case of environmental suppressed $g_{a\gamma}$ (dashed curve). The expected background is represented by horizontal dashed line.}
\label{fig:nev}
\end{figure}

 From Fig.~\ref{fig:alp_flux}, it is also clear that for a coupling $g_{\rm PSR}$ the environmentally suppressed
 solar ALP flux is peaked at low energies, below the CAST threshold for solar axion searches
 ($E_{\rm th}=2$~keV). In this region it would exceed the standard flux.
 At this regard it is intriguing that CAST has also performed 
 a search for soft X-ray photons in the energy range from 200 eV to 10 keV~\cite{Anastassopoulos:2018kcs}  with  GridPix detector. 
 This was relevant to study possible
 conversions of solar chameleons~\cite{Brax:2010xq,Anastassopoulos:2015yda}. 
 The presence of these data allows us to compare the expected signal from a low-energy ALP flux with 
 the existing data. In particular, they used as energy window for  low-energy flux searches 
 $E\in[0.2;2]$~keV  
 where they reach a background rate $10^{-3}-10^{-4}$/keV/cm$^{2}$/s~\cite{Anastassopoulos:2018kcs}.
 One can calculate the expected X-ray flux from low-energy ALPs, multiplying the solar ALP flux by
 the conversion probability in the CAST magnet, which for low-mass ALPs reads~\cite{Andriamonje:2007ew}
 %..................
 \begin{equation}
 P_{a\gamma} \simeq 1.7 \times 10^{-17} g_{10}^{2} \left(\frac{B L}{ 9  \textrm{T} \times 9.26  \textrm{m}} \right)^{2} \,\ ,
 \end{equation}
 %.......................
 where $B$ is the CAST magnetic field and $L$ is the magnet length.
The expected event rate in the energy-window $E\in[0.2;2]$~keV is shown in Fig.~\ref{fig:nev}
for the standard case (continuous curve) 
and the case of environmentally suppressed $g_{a\gamma}$ (dashed curve). The expected background is represented by the horizontal dashed line.
It results that in this energy window the expected event rate in the case of an environmentally suppressed $g_{a\gamma}$ would be 
between three and four orders of magnitude below the estimated background.  Therefore, a significant background reduction
would be mandatory in order to probe this scenario.
 
%%%%%%%%%%
%%%%%%%%%%%%
 
 %%%%%%%%%%%%%%%%%%%%
 \subsection{Globular Clusters}
 \label{sec:GC}
 %%%%%%%%%%%%%%%%%%%%% 
 
 Aside from the relaxation of the CAST bound, the environmental suppression of the $g_{a\gamma}$ coupling would significantly 
 weaken also the other astrophysical ALP constraints, specifically the globular cluster and supernova SN 1987A bounds.
 
 In the context of globular clusters,
 a particularly sensitive observable is the $R$ parameter, 
defined as the number ratio of horizontal branch (HB) stars to red giants branch (RGB)  stars in a globular cluster, 
$R= N_{\rm HB}/R_{\rm RGB}$.
The $R$ parameter is known to be particularly efficient in constraining the 
ALP-photon coupling~\cite{Raffelt:1987yu,Ayala:2014pea}.
For low-mass ALPs, the most relevant ALP production mechanism induced by the photon coupling 
is the Primakoff process.
%like in the case of the Sun. 
This is considerably more efficient in HB than in RGB stars,
since in the latter it is suppressed by electron degeneracy effects and by a larger plasma frequency. 
Therefore, in the presence of ALPs with a sizable  $g_{a\gamma}$   
one would expect a significant reduction of the HB lifetime, causing a reduction of the $R$ parameter.
In Ref.~\cite{Ayala:2014pea}, it was shown that consistency with the $R$ parameter observed in 39 Galactic globular clusters required
 $g_{10} \leq 0.66$.
 
\begin{figure}[t!]
\center
\includegraphics[width=0.6\columnwidth]{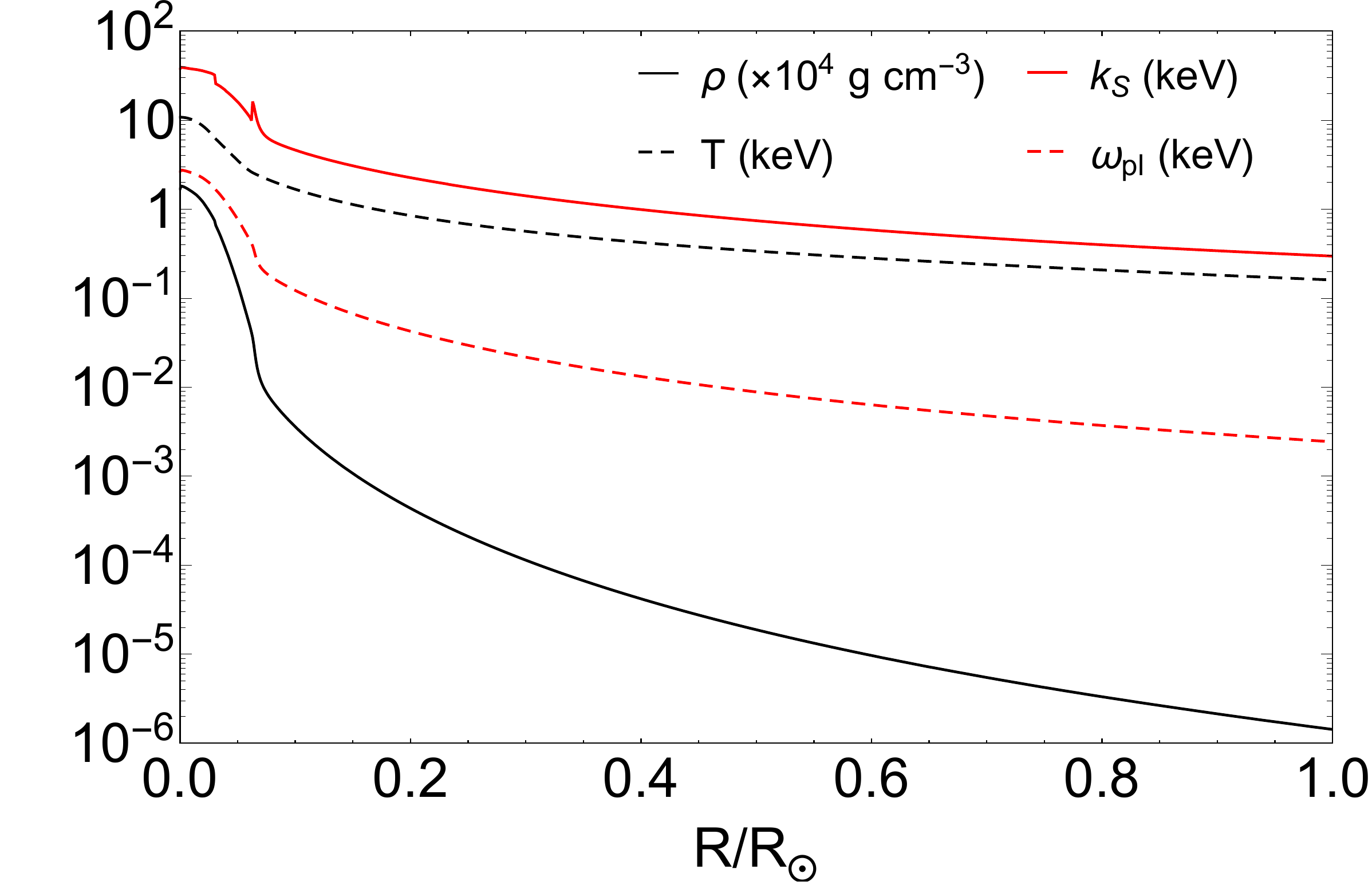}
\caption{Radial evolution of the 
environmental parameters in the HB stars.}
\label{fig:HB_model}
\end{figure}

 The axion emission rate (energy per mass per time) via the Primakoff process is given by the expression \cite{Anastassopoulos:2017ftl,Lucente:2020whw}
%.........................
\begin{equation}
\varepsilon_a = \frac{2}{\rho} \int \frac{dp \,p^2}{2 \pi^2} \Gamma_{\gamma\to a}\, E\, f(E) \,\ ,
\end{equation}
%...............................
where the factor 2 comes from the photon degrees of freedom,
$ \rho $ is the local density, 
 $f(E) = (e^{E/T}-1)^{-1}$ is the Bose-Einstein distribution, and 
$\Gamma_{\gamma \to a}$ is the photon-ALP transition rate, given in Eq.~(\ref{eq:primakoff}). 
The ALP luminosity is obtained integrating the emissivity over a stellar profile
%.................
\begin{equation}
L_a = 4 \pi \int  \rho \varepsilon_a r^2 dr \,\ .
\end{equation}
The radial evolution of the 
environmental parameters within the helium-rich core of a typical HB stellar model 
is shown in Fig.~\ref{fig:HB_model} (see~\cite{Ayala:2014pea} for details). 

\begin{figure}[t!]
\center
\includegraphics[width=0.6\columnwidth]{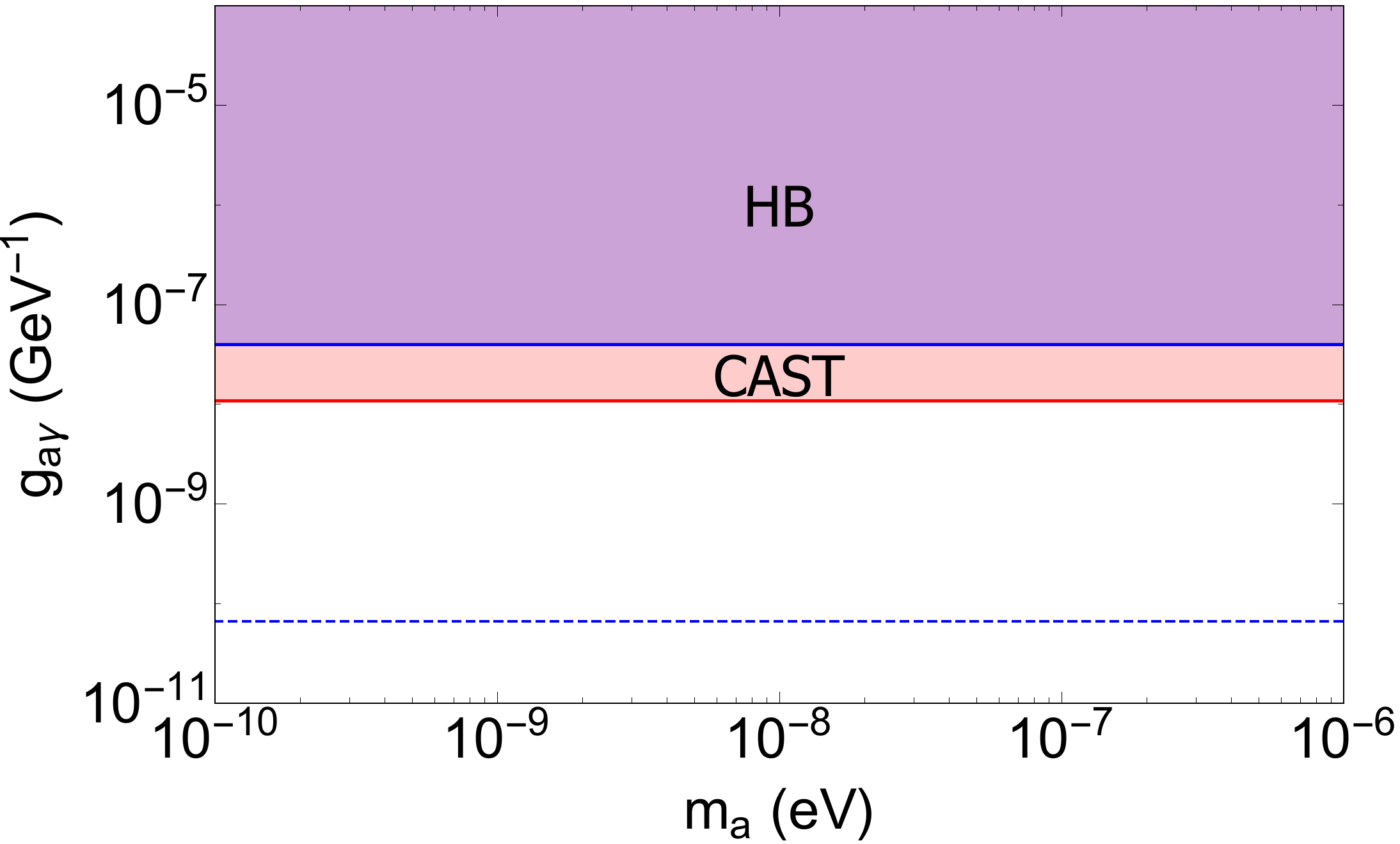}
\caption{Relaxed ALP bounds from CAST and HB stars in a model with environmental suppression of photon-ALP coupling. The blue dashed line is
the standard HB bound}.
\label{fig:newBound}
\end{figure}

The limiting value  $g_{a\gamma} = 0.66  \times 10^{-10}$  {GeV}$^{-1}$ 
 corresponds to an energy loss
 $ \varepsilon_a  \lesssim 38 \,\  \textrm{erg} \,\ \textrm{g}^{-1} \textrm{s}^{-1}$, and to 
 an integrated ALP luminosity $L_a \lesssim 10^{34}\,\  \textrm{erg} \,\ \textrm{s}^{-1}$.
 For definitiveness, we assume that the environmental suppression in the HB star depends
 on the density $\rho$ and, in order to be consistent with the Sun case, we fix the 
 critical density at $\rho_c= 1.3$ g\,cm$^{-3}$.   From a comparison with Fig.~\ref{fig:newBound}
it results that with this choice one would lose most of the ALP emissivity. 
 Imposing the luminosity bounds, one finds
 %.......................
 \begin{equation}
 g_{10} < 4 \times 10^{2} \,\ ,
 \end{equation}
 %...........................
 implying a significant relaxation of the HB bound with respect to the standard case.
 A similar relaxation would be found assuming a suppression of the coupling depending on other 
 environmental parameters.

%%%%%%%%%%
%%%%%%%%%%%%

 %%%%%%%%%%%%%%%%%%%%
 \subsection{SN 1987A}
  \label{sec:SN}
 %%%%%%%%%%%%%%%%%%%%% 
 
 In a core-collapse supernova, ALPs would be emitted via the Primakoff process, and eventually convert into gamma rays in the magnetic field of the Milky Way. The lack of a gamma-ray signal in the GRS instrument of the SMM satellite in coincidence with the observation of the neutrinos emitted from SN1987A therefore provided a strong bound on their coupling to photons. Notably for $m_a < 4 \times 10^{-10}$~eV the most recent analysis finds
 $g_{10} < 5.3 \times 10^{-2}$~\cite{Payez:2014xsa}. 
The environmental suppression of $g_{a\gamma}$ in a SN matter, being calibrated on the Sun conditions, would be dramatic. Indeed, it would imply that
the ALP-photon coupling would be vanishing in the core where typical densities would be $\rho \sim 10^{14}$ g\,cm$^{-3}$
and temperature $T\sim 30$~MeV. Therefore, the ALP would be emitted only from the outer layers of the star with a strongly reduced flux and an energy outside the 
band of the SMM satellite. Therefore, in this model the ALP bound would practically disappear.
A similar suppression would apply also to bounds from extra-galactic core-collapse supernovae~\cite{Meyer:2020vzy}.

 %%%%%%%%%%
%%%%%%%%%%%%
 
%%%%%%%%%%%%%%
\subsection{Signal in ALPS II}
\label{sec:ALPSII}
%%%%%%%%%%%%%   

An ALP-photon coupling $g_{a\gamma}$ of the size required to explain the spectral modulation observed in Galactic PSR gamma-ray spectra, $g_{\rm PSR} = 1.97 \times 10^{-10}$\,GeV$^{-1}$, would lead to a spectacular signal rate in the pure laboratory 
experiment ALPS II \cite{Bahre:2013ywa} which is expected to start data taking in 2021. ALPS II exploits the light-shining-through-walls technique (for a review, see Ref. \cite{Redondo:2010dp}) to produce and detect ALPs. It consists of two aligned and matched optical cavities
separated by a wall. Both cavities are placed in a transverse dipole magnetic  
field of strength $B=5.3$\,T and length $L= 12\times 8.83$\,m. The ALP generation cavity is powered by an infrared laser (wave length $\lambda =1064$\,nm, power $P_{\rm prim}=30$\,W) and designed to have power build up 
$\beta_{\rm g}=5\times 10^3$. ALPs, if they exist, can be produced via ALP-photon conversion in the magnetic 
field of the ALP generation cavity and propagate through the wall into the photon regeneration cavity (power build up $\beta_{\rm r}=4\times 10^4$) behind the wall, where they may convert 
again into photons of energy $\omega$. Since the ALP-photon conversions happen in vacuum, the ALP will not experience any environmental and thus no eventual suppression of the photon coupling. In fact, for small ALP masses,  
$m_a\ll   2 \sqrt{{\omega}/{L}}\simeq 1.5\times 10^{-4}$\,eV, 
where $\omega = 2\pi/\lambda=1.16$\,eV is the laser photon energy, the expected rate of regenerated photons \cite{Hoogeveen:1990vq,Mueller:2009wt} at ALPS II,
\begin{eqnarray}
\frac{\Delta N_r}{\Delta t} &\simeq&     \frac{P_{\rm prim}}{\omega}  \beta_{\rm g} \beta_{\rm r} \left[ \frac{1}{4} g_{a\gamma}^2 B^2 L^2\right]^2 \nonumber \\
& \simeq &
0.29\,{\rm Hz} 
\left[ \frac{g_{a\gamma}}{1.97\times 10^{-10}\,{\rm GeV^{-1}}}\right]^4
\,,
\end{eqnarray}
is four orders of magnitudes larger than the expected background rate,  
implying a clear ALP discovery.

\section{Conclusions}
\label{sec:conclusions}

Astrophysics offers a valuable tool to probe very light ALPs. 
An ALPs flux can be produced in stellar cores via the Primakoff process. 
One can probe it indirectly requiring that it does not contribute to an excessive energy loss in stellar systems such as helium burning stars or SN 1987A.  
In the case of the Sun, the ALPs flux is expected to be so intense that one can directly search for it through ALP-photon conversions
in a laboratory magnetic field, as done by the CAST helioscope experiment. 
These arguments at the moment lead to stringent bounds on the ALP-photon coupling $g_{a\gamma}$. 

Other searches for ALPs look for signatures of ALP-photon conversions in cosmic magnetic fields.
In fact, high-energy gamma-ray observations also provide stringent bounds on the ALP-photon coupling for ultralight particles. 
In this context, intriguing hints recently emerged. Notably, it has been claimed that  the spectral modulation observed in gamma rays
from Galactic PSRs and supernova remnants can be due to  conversions of photons into ultra-light ALPs in large scale Galactic  magnetic fields. 
These hints appear to be in tension with the stellar bounds. 
Here, we have shown how they can be reconciled with the known experimental and astrophysical bounds, assuming that the 
ALP-photon coupling has an environmental dependence that suppresses it in the dense stellar plasma, leaving it unaffected in the low-density
Galactic environment. 
We have discussed the phenomenological implications of this scenario, and
shown how the CAST bound and the constraints from helium burning stars and SN 1987A would be relaxed under this assumption, 
relieving the tension with the PSR claim.
Furthermore, this scenario is directly testable  in the light-shining-through-the-wall experiment ALPS II,
which is expected to be operative in a year or so.

\medskip

\section*{Acknowlegments}
The work of P.C. and 
A.M. is partially supported by the Italian Istituto Nazionale di Fisica Nucleare (INFN) through the ``Theoretical Astroparticle Physics'' project
and by the research grant number 2017W4HA7S
``NAT-NET: Neutrino and Astroparticle Theory Network'' under the program
PRIN 2017 funded by the Italian Ministero dell'Universit\`a e della
Ricerca (MUR). D.H., A.R. acknowledge support and A.S. is funded by the Deutsche Forschungsgemeinschaft (DFG, German Research Foundation) under Germany’s Excellence Strategy – EXC 2121 \textit{Quantum Universe} – 390833306.
G.A.P. acknowledges support from the Charpak Lab fellowship for carrying out a 3-month internship at LAPTh, Annecy.

\newpage

\appendix
\section{Galactic PSR sample}
\label{app:psr_sample}
We report here some relevant information about the PSR sample used for the present analysis.

In Tab.~\ref{Tab:psr}, we quote PSR name, position, distance estimate from the latest ATNF catalog version, and FWHM of the distance PDF.

\begin{figure*}
\centering
\includegraphics[scale=0.55]{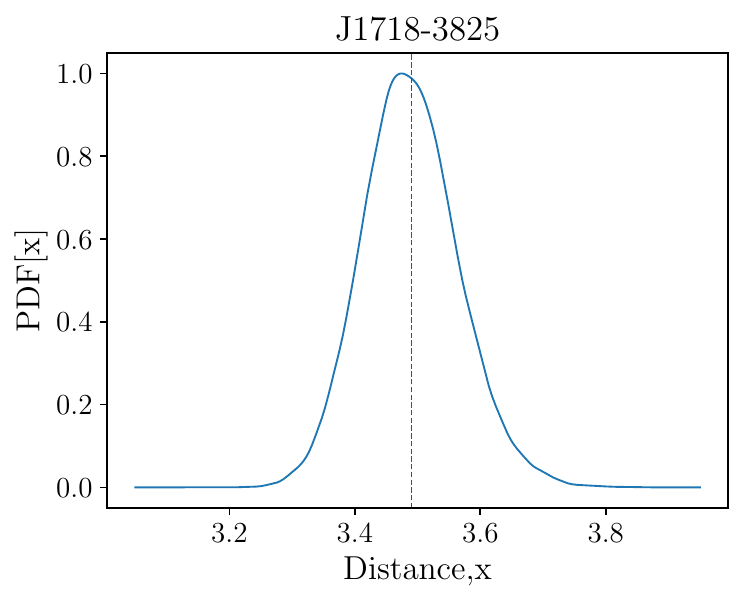}
\includegraphics[scale=0.55]{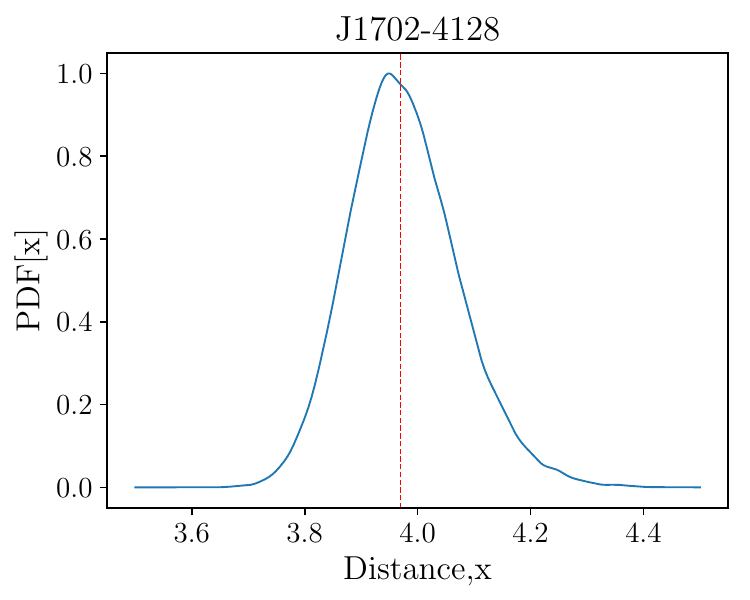} \\
\includegraphics[scale=0.55]{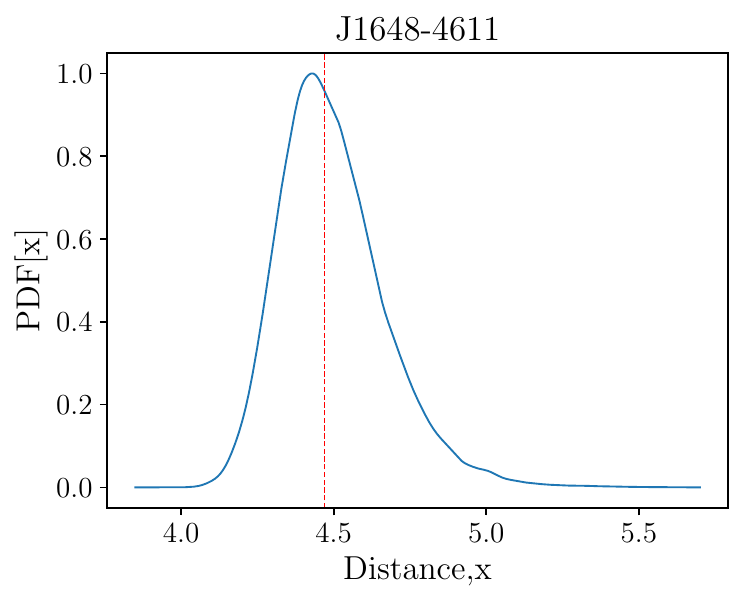}
\includegraphics[scale=0.55]{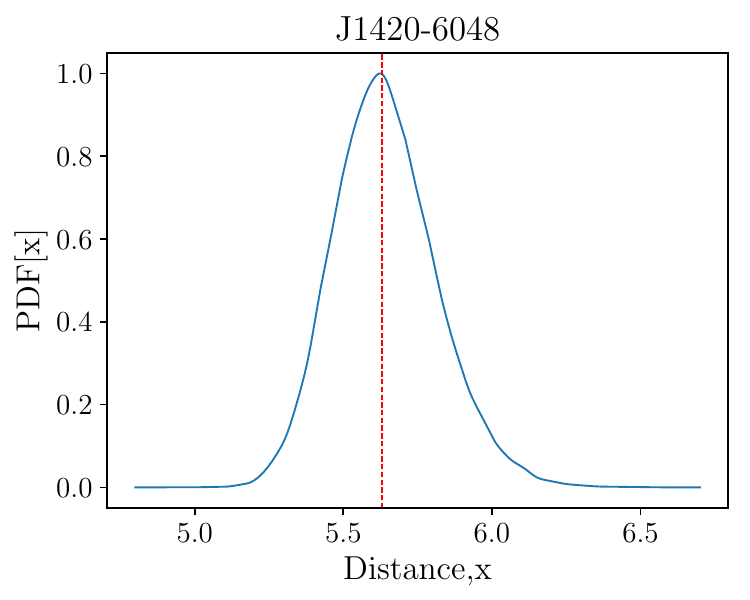}\\
\includegraphics[scale=0.55]{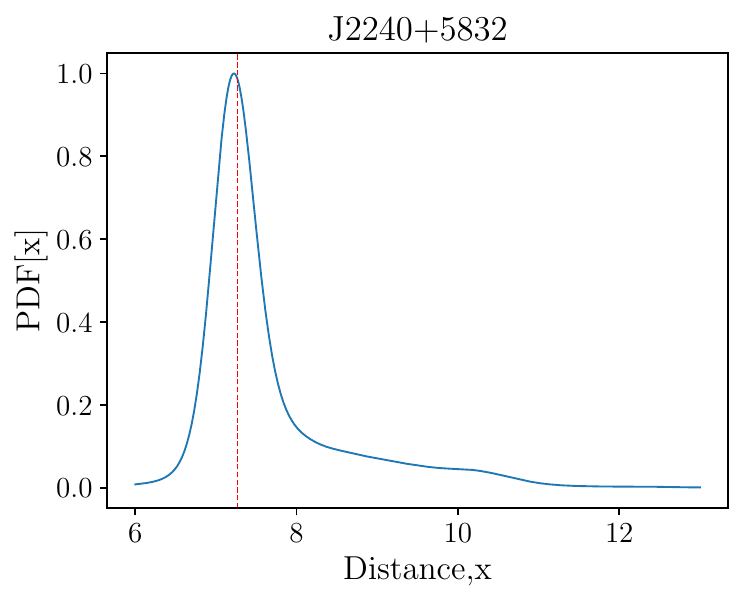}
\includegraphics[scale=0.55]{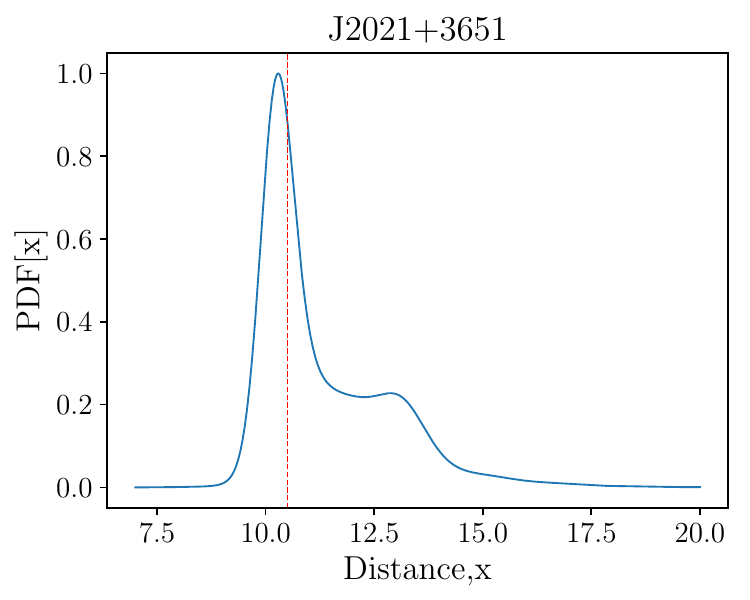}
\label{fig:psr_pdf}
\caption{PSR's distance PDF (blue), together with the distance estimate from the ATNF catalog (red line).}
\end{figure*}

The PDF are obtained following the numerical implementation of~\cite{Bartels:2018xom}.
Technically, a discrete PDF is obtained by sampling over a large number of distances and then interpolated to obtain a continuous function, which is then used for subsequent analysis. 
We normalize the PDF to be 1 at the peak.
The distance PDFs for all PSRs are shown in Fig.~\ref{fig:psr_pdf}.

\begin{table}[ht]
\caption{PSR sample used in the present analysis, together with Galactic coordinates positions, distance estimates from the latest version of the ATNF catalog, and FWHM of the distance PDF.}
\vspace{0.5cm}
\centering
\renewcommand{\arraystretch}{1.4}
\setlength{\tabcolsep}{15pt}
	\begin{tabular}{|c|c|c|c|c|}
 	\hline
 	PSR Name & $l$ & $b$ & d [kpc] & FWHM [kpc]\\
 	\hline
 	J1718-3825 & 348.951 & -0.432 &   3.49 & 0.17\\
 	J1702-4128 & 344.744 & 0.123 & 3.9 & 0.21\\
 	J1648-4611 & 339.438 & -0.794 & 4.47 & 0.36\\
 	J1420-6048 & 313.541 & 0.227 & 5.63 & 0.39\\
 	J2240+5832 & 106.566 & -0.111 & 7.27 & 0.66\\
 	J2021+3651 & 75.222 & 0.111 & 10.51 & 1.02\\
 	\hline
	\end{tabular}
	\label{Tab:psr}
\end{table}

\section{PSR signal region: PSR spectra and statistical framework}
\label{app:chi2_psr}
We here report the details of extraction of PSR spectra and of the statistical framework used to find the best-fit and define the PSR signal region.
%

% details about extraction of PSR spectral points
As said in the main text, we use the gamma-ray spectra of the six bright Galactic PSRs as they were previously published in~\cite{Majumdar:2018sbv}.

In~\cite{Majumdar:2018sbv}, we used gamma-ray spectra of the six bright galactic pulsars derived with 9 years of {\it Fermi}-LAT dataset spanning from August 2008 to August 2017. The dataset was based on the Pass8 SOURCE photon reconstruction in the energy range from 100 MeV to 300 GeV. We focused only on the  SOURCE class events with a maximum zenith angle of $90^{\circ}$ to exclude the contamination from the Earth limb emission. The LAT data analysis was performed
with the \texttt{Enrico} python package which is  based on the \texttt{Fermi Science Tools} (\cite{Sanchez:2013ixa}). Data within  $10^{\circ}$ from the source position (center of the region of interest, ROI)
 were binned into 8 energy bins per decade with a spatial
bin size of $0.1^{\circ}$ for the analysis. We included the templates for the Galactic  (\texttt{gll\_iem\_v06.fits}) and the isotropic (\texttt{iso\_P8R2\_SOURCE\_V6\_v06.txt}) diffuse components.  All point sources listed in the third {\it Fermi}-LAT source catalog \cite{Fermi-LAT:2015bhf} within $15^{\circ}$
from the ROI center of the pulsar were included in the analysis. In the spectral analysis, pulsar spectra were modeled with a power law with exponential cutoff given by:
\begin{equation}
 \frac{dN}{dE} =N_{\circ} \left(\frac{E}{E_{\circ}} \right)^{-\Gamma_1} \exp \left[-\left(\frac{E}{E_{\rm cut}} \right)^{\Gamma_{2}}\right]
 \end{equation}
 
where, $\rm N_{\circ}$ is the normalization factor at the scale energy ($\rm E_{\circ}$), $\Gamma_1,\Gamma_2$ are the photon-indices, and $\rm E_{cut}$ is the cutoff energy.

In the likelihood fit, the spectral parameter of all the sources within $3^{\circ}$ from the ROI centre, as well as the diffuse emission components, were left free to vary while the parameters for other sources kept fixed. The gamma-ray flux was obtained by performing a binned likelihood analysis in each energy bin and the resulting {\it Fermi}-LAT spectral energy distribution was then used to study  photon-ALPs oscillations, as done also in the present work.  

We here notice that performing an on/off phase-resolved analysis of the PSR spectrum can help more robustly constraining the PSR emission and controlling potential systematic uncertainties related to foreground/background mis-modelling. 
Indeed, an iterative fit, which exploits the off-pulse data to constrain diffuse emission and nearby point-sources independently of the PSR, can break potential degeneracies between the fit parameters present when an analysis over the entire phase period is performed.

% statistical analysis
Following~\cite{Majumdar:2018sbv}, we run fits of the PSRs spectra under different hypotheses, $H_0$, $H_1$, and $H_2$, as sketched in  
Table.~\ref{tabl:Hypothesis}. 
\begin{table}[ht]
\caption{Description of tested Hypotheses}
\label{tabl:Hypothesis}
\vspace{0.5cm}
\centering
\renewcommand{\arraystretch}{1.4}
\setlength{\tabcolsep}{15pt}

	\begin{tabular}{|c|c|}
 	\hline
 	Hypothesis & Assumptions\\
 	\hline
 	$\textrm{H}_{0}$ & No ALP Fit \\
 	$\textrm{H}_{1}$ & Fitting with ALP with $\mathrm{m_{a}}$ and $\mathrm{g_{a\gamma}}$ left free for each pulsar \\
 	$\textrm{H}_{2}$ & Fitting with ALP with $\mathrm{m_{a}}$ and $\mathrm{g_{a\gamma}}$ globally fit \\
 	\hline
	\end{tabular}
	\label{Tab:psr}
\end{table}
\newpage

While reproducing the results of~\cite{Majumdar:2018sbv}, we found some inconsistencies between the numerical implementation of the spectral fits and what was written in the paper. In particular, there were inconsistencies in the spectral parameter $\Gamma_{2}$ used in the code and the value quoted in the text. We have now fixed consistently $\Gamma_{2}=0.54$ for all PSRs, as written in ~\cite{Majumdar:2018sbv}. Additionally, there were some errors in the previous implementation of Galactic magnetic field, which has since been corrected.\footnote{To model the Galactic magnetic field, we now use the latest version of the code from \url{https://github.com/me-manu/gammaALPs}.}
After including these corrections to the analysis in~\cite{Majumdar:2018sbv}, we obtain the final results in Table  \ref{tabl:CompareH}

\begin{table}[ht]
\caption{Comparison of best-fit $\chi^{2}$ between the three hypotheses tested.}
\label{tabl:CompareH}
\vspace{0.5cm}
\centering
\renewcommand{\arraystretch}{1.4}
\setlength{\tabcolsep}{15pt}

	\begin{tabular}{|c|c|c|c|c|}
 	\hline
 	PSR Name & $\chi^{2}  \mathrm{(dof) \hspace{5pt} H_{0}}$ & $\chi^{2}  \mathrm{(dof) \hspace{5pt} H_{1}}$ & $\chi^{2}  \mathrm{(dof) \hspace{5pt} H_{0}}$\\
 	\hline
 	J1718-3825 & 28.61(15) & 11.25(13) &  28.39(15) \\
 	J1702-4128 & 10.08(8) & 2.08(6) & 9.63(8)\\
 	J1648-4611 & 26.64(14) & 17.19(12) & 24.15(14) \\
 	J1420-6048 & 22.96(15) & 18.44(13) & 20.13(15)\\
 	J2240+5832 & 13.58(11) & 4.61(9) & 9.78(11)\\
 	J2021+3651 & 44.71(14) & 17.35(12) & 25.48(14)\\
 	\hline
 	Combined & 146.58(77) & 70.92(65) & 117.56(75)\\
 	\hline
	\end{tabular}
	\label{Tab:psr}
\end{table}

Since we do not know the distribution of the TS, we perform Monte Carlo simulations of the PSRs spectra using \texttt{fermipy}, and based on the null hypothesis fits to the six pulsars.
In particular, we use \texttt{fermipy} version $0.19.0$.
We run \texttt{gta.simulate\underline{\,\,\,}roi} to generate the simulated counts cube of the ROI from the best-fit model under the null hypothesis derived here.
Poisson noise is added to the simulated counts cube and 50 simulations are created for each PSR, in order to study the impact of statistical fluctuations. 
After this step, the newly created counts files are processed by an analysis pipeline for the spectral reconstruction (running \texttt{gta.optimze} and \texttt{gta.fit}), which closely follow the procedure from Ref.~\cite{Majumdar:2018sbv} described above.

We therefore obtain 50 simulated spectral data sets for each pulsar. These simulations are used to derive the TS distribution analogously to~\cite{TheFermi-LAT:2016zue}. 
To obtain TS distribution, we find $\mathrm{TS_{\rm max}}$ by obtaining the maximum $\mathrm{TS}$ for each data set. Since we have only 50 simulated data sets for each pulsar, a direct analysis can give us 50 sample sets of six pulsars to obtain the TS distribution. However, we can improve our statistics by applying the bootstrap method. Here, we randomly choose 6 data sets, by picking a random data set from the 50 simulated data sets of a given pulsar and repeating it for all pulsars. For this new random set, we can obtain the TS values by fitting it with both $\mathrm{H_{0}}$ and $\mathrm{H_{2}}$ hypotheses. We can repeat this procedure, and arbitrarily increase the sampling of the TS distribution. We chose $10^{4}$ of such realizations.

\begin{figure}
\centering
\includegraphics[scale=0.4]{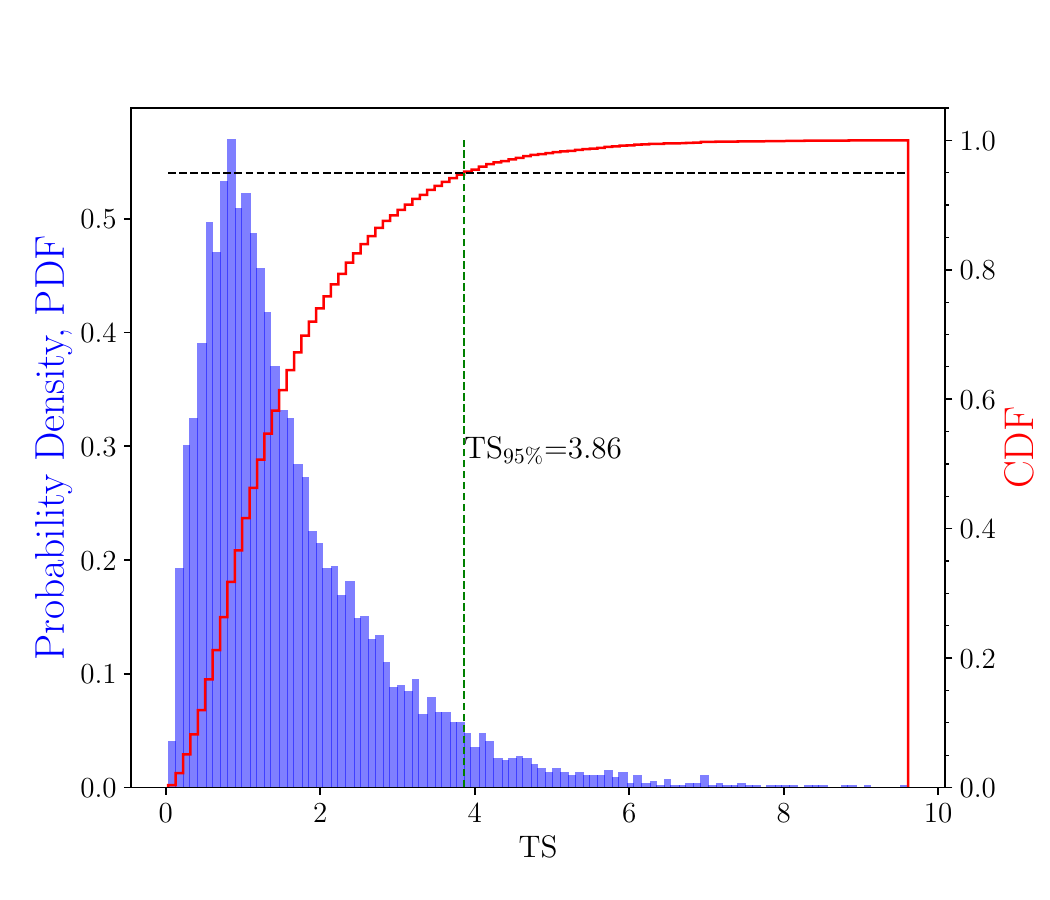}
\includegraphics[scale=0.4]{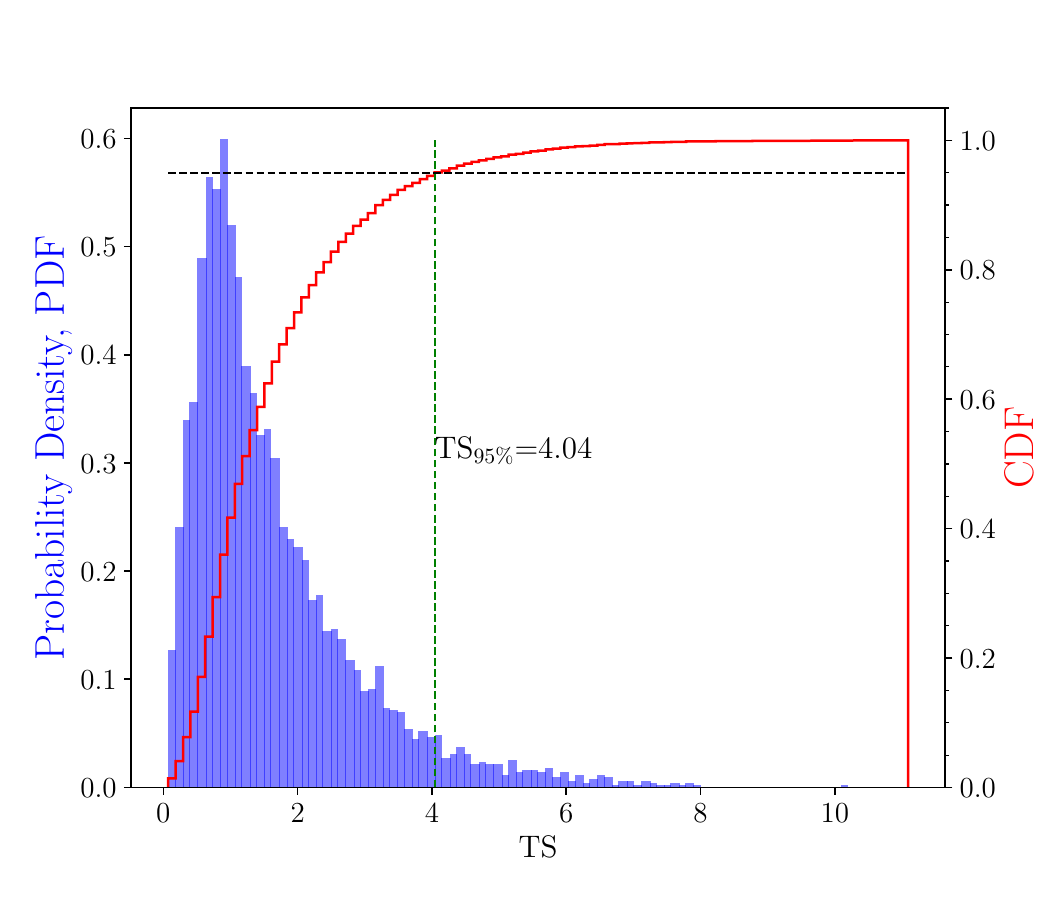}\\
\caption{\textit{Left panel}: TS distribution for simulated data. \textit{Right panel}: TS distribution for simulated data when profiling over distance uncertainty of each PSR.}
\label{fig:TS}
\end{figure}

The results of such an analysis are shown in left panel of Fig.~\ref{fig:TS}. This corresponds to fits without profiling over distance or Galactic magnetic field uncertainties. One can immediately see that, while the ALP hypothesis does improve the fits, the improvement is not significant for the simulated data sets. Therefore, we can expect already that repeating the analysis while profiling over Galactic magnetic field and distance uncertainties would not change the TS distribution significantly since they would only slightly modify the ALP fit, but not alter it significantly as can been seen from Fig.~\ref{fig:PSR_region}. To verify this we repeated the full analysis for the case of profiling over distance uncertainties. The results, shown in right panel of Fig. \ref{fig:TS}, confirms our expectation. Since doing a similar analysis for the case of profiling over Galactic magnetic field is computationally quite intensive, we adopt $\mathrm{TS_{95\%}}=3.86$ from Fig. \ref{fig:TS} to draw the contours for all three scenarios - as justified by the above argument. 

While one can, in principle, calculate the significance of the signal directly using TS distribution, we note that the $\mathrm{TS_{\rm max}}$ values in Fig. \ref{fig:PSR_region} is much larger than the maximum in the TS distribution we obtained from the simulation. Thus, one cannot obtain any reliable result using the TS distribution, and hence we resort to F-test. We follow the same procedure as described in \cite{Majumdar:2018sbv}. The results are compiled in Table \ref{tabl:ftest}.

\begin{table}[ht]
\vspace{0.5cm}
\centering
\renewcommand{\arraystretch}{1.4}
\setlength{\tabcolsep}{15pt}

	\begin{tabular}{|c|c|c|c|c|}
 	\hline
 	Hypothesis & $\mathrm{\chi^{2}}$ & dof & Significance ($\mathrm{H_{i}}/\mathrm{H_{0}}$)\\
 	\hline
 	$\mathrm{H_{0}}$ & 146.58 & 77 &  - \\
 	$\mathrm{H_{1}}$ & 70.92 & 65 & 4.73 $\sigma$\\
 	$\mathrm{H_{2}}$ & 117.56 & 75 & 3.47 $\sigma$ \\
 	$\mathrm{H_{2d}}$ & 109.40 & 69 & 2.45 $\sigma$\\
 	$\mathrm{H_{2B}}$ & 104.95 & 70 & 3.07$\sigma$\\
 	\hline
	\end{tabular}
\caption{Significance calculation using F-test: $\mathrm{H_{2d}}$ is $\mathrm{H_{2}}$ while profiling over uncertainties in pulsar distance and $\mathrm{H_{2B}}$ is $\mathrm{H_{2}}$ while profiling over uncertainties in Galactic magnetic field}
\label{tabl:ftest}
\end{table}

%\bibliographystyle{plain}
%\bibliography{pulsar_bib}

\end{document}